\documentclass[acmsmall]{acmart}
\AtBeginDocument{%
  \providecommand\BibTeX{{%
    \normalfont B\kern-0.5em{\scshape i\kern-0.25em b}\kern-0.8em\TeX}}}

\setcopyright{none}
\renewcommand\footnotetextcopyrightpermission[1]{}

\setcopyright{none}
\renewcommand\footnotetextcopyrightpermission[1]{}

\makeatletter
\def\@copyrightspace{\relax}
\makeatother

\begin{document}

\title{A complete discussion on fully reconfigurable, digital, scalable, graph
and sparsity-aware near-memory accelerator for graph neural
networks}

\author{Siddhartha Raman Sundara Raman}
\author{Lizy John}
\author{Jaydeep P. Kulkarni}
\affiliation{\institution{The University of Texas At Austin} \country{USA}  \\This is an extension of the paper published in ACM TACO 2024}

\renewcommand{\shortauthors}{Siddhartha Raman, et al.}

\begin{abstract}
Graph neural networks (GNNs) have gained significant interest for applications such as citation network analysis and drug discovery due to their ability to apply machine learning techniques on graph-structured data. GNNs typically employ a two-stage execution pipeline consisting of combination and aggregation kernels. The combination stage performs data-intensive convolution operations with relatively regular memory access patterns, whereas the aggregation stage operates on sparse graph data with highly irregular accesses. These heterogeneous memory behaviors make conventional CPU- and GPU-based execution energy inefficient due to substantial data movement overheads.

Existing accelerators attempt to mitigate these challenges using specialized architectures and processing-in-memory (PIM) techniques. However, prior approaches often suffer from scalability limitations, area overheads, restricted parallelism, and energy inefficiencies associated with analog compute and dedicated accelerator structures. 

This paper presents NEM-GNN, a scalable DAC/ADC-less processing-in-memory architecture for graph neural network acceleration. The proposed design introduces early compute termination mechanisms, pre-computation using reconfigurable system-on-chip components, and graph- and sparsity-aware near-memory aggregation using a compute-as-soon-as-ready (CAR) and broadcast-based execution model. Experimental results demonstrate that NEM-GNN achieves approximately 80--230$\times$ higher performance, 80--300$\times$ higher throughput, 850--1134$\times$ better energy efficiency, and 7--8$\times$ higher compute density compared to prior state-of-the-art approaches.

\end{abstract}


%\ccsdesc[300]{Do Not Use This Code~Generate the Correct Terms for Your Paper}
%\ccsdesc{Do Not Use This Code~Generate the Correct Terms for Your Paper}
\begin{comment}
\ccsdesc[100]{Do Not Use This Code~Generate the Correct Terms for Your Paper}
\end{comment}
\begin{comment}
\received{20 February 2007}
\received[revised]{12 March 2009}
\received[accepted]{5 June 2009}
\end{comment}
\maketitle
\pagestyle{plain}
\thispagestyle{plain}

\makeatletter
\def\@oddfoot{}
\def\@evenfoot{}
\makeatother

\section{Introduction}
Over the past decade, there has been widespread utilization of deep learning models such as convolutional neural networks (CNN), and recommendation networks in diverse fields like image and video processing. These models primarily operate within the realm of Euclidean data, wherein data inputs conform to a structured and precisely defined n-dimensional space. However, these well-established models exhibit inefficiency when confronted with domains like citation networks, molecular chemistry, and electric grids \cite{Citation_network}\cite{ReFlip}, characterized by non-Euclidean data structures such as graphs and manifold geometries (3D surface structures). In these cases, the data inputs deviate from structured paradigms, and the pursuit to encapsulate them within a strictly defined n-dimensional space leads to loss of valuable information. Consequently, Geometric Deep Learning (GDL) \cite{GDL} has emerged as a solution tailored to the idiosyncrasies of non-Euclidean data relationships. One of the approaches geared towards processing graph-based data hinges on the utilization of Graph Neural Network (GNN) models. These models are adept at handling graph-based inputs, facilitating the classification of nodes in a graph into distinct categorical groups. %Thus, Geometric deep learning (GDL) \cite{GDL} is developed to handle the relationships among non-Euclidean data inputs. 

%GDL approach for processing graphs uses graph neural network (GNN) \cite{GCN} models, which take graphs as inputs and classify the graph nodes into different categories. %An example of GNN-based processing for citation networks involves the classification of published articles in a repository by representing articles as graph nodes and directed edges between the nodes indicating an article cited another \cite{GCN_cite}. 
\par Graph Neural Networks (GNNs) employ two fundamental mechanisms:
(i) A combination kernel, akin to the convolution process in CNNs, is harnessed to encode the information within nodes. (ii) Aggregation kernels are then utilized to encode the information in edges, which helps to comprehend the relationship between graph nodes. Regarding computations, the operations associated with the combination kernel are notably data-intensive, exhibiting regular memory access patterns. On the other hand, operations linked to the aggregation kernel are characterized by sparsity and irregularity. The mixed data pattern is a major limiter behind using CPU/GPU for GNN compute  \cite{Hygcn}. Various accelerator designs have been proposed to tackle this particular concern. 
\par  These accelerators can broadly be classified into traditional Von-Neumann-based accelerators/ processing-in/near memory designs. The proposed designs (both Von-Neumann/PIM) are dedicated domain-specific accelerators, that require periodic interaction with the host, resulting in energy overhead. The existing Von-Neumann-based accelerator designs like AWB-GCN\cite{AWB-GCN}, HyGCN\cite{Hygcn} incur data movement cost for transferring data from the processor to memory. Since the combination kernel uses data-intensive operations, requiring periodic data movement costs, this architecture is energy inefficient. The state-of-the-art processing in/near-memory (PIM/PNM) architectures for solving traditional graph algorithms, and GNNs, like ReFLIP \cite{ReFlip} exhibit reduction in data movement by enabling computations within memory \cite{IGZOCIM}. These architectures use crossbar Resistive Random Access Memory (ReRAM) arrays, Digital-to-Analog converters (DAC), Analog-to-Digital converters (ADC). The problems specifically with ReFLIP are: (i) This is a \textbf{dedicated accelerator} requiring periodic host-accelerator interaction, leading to energy inefficiency. (ii)  The presence of DACs/ADCs renders the architecture susceptible to process variations,  causing a decline in \textbf{accuracy}. (iii) Exponential increase in storage requirement with increased bit-precision/resolution for GNN compute, incurring \textbf{scalability} challenges. (iv) Aggregation is performed only upon the completion of combination operations, lacking any overlap between them, limiting \textbf{performance} (v) Sparse in-memory aggregation leads to \textbf{compute density} challenges. The major contributions are:
\begin{itemize}
     \item \textbf{Re-configurability}: SOC components like L1/L2 cache are reconfigured to realize NEM-GNN without requiring memory array modifications/dedicated accelerator arrangement, thereby improving energy of the design. 
     \item \textbf{DAC/ADC-less}: Bit-serial PIM architectures for combination (NEM-C1, NEM-C2, NEM-C3) with early termination of compute, pre-compute strategies without DAC/ADC, achieving highly accurate compute.
     \item \textbf{Scalability}: NEM-GNN is scalable to high bit-precision compute without requiring exponential storage for the compute array.
    \item \textbf{Graph aware}:
    Near-memory aggregation optimized specifically with "Compute-as-soon-as-ready" (CAR), "broadcast" approaches to hide aggregation latency, by overlapping combination and aggregation completely. 
    \item \textbf{Sparsity aware}: Leveraging \textbf{sparsity} in kernels to minimize the unnecessary compute and utilizing the high-throughput of the design to improve compute density further.   
    \item NEM-GNN outperforms ReFLIP by $\sim$80-230x, $\sim$80-300x, $\sim$850-1134x and $\sim$7-8x in terms of performance, throughput, energy efficiency and compute density \cite{NEM_GNN}.
\end{itemize} 

\begin{figure*}[t!]
\centering
\includegraphics[width=\linewidth]{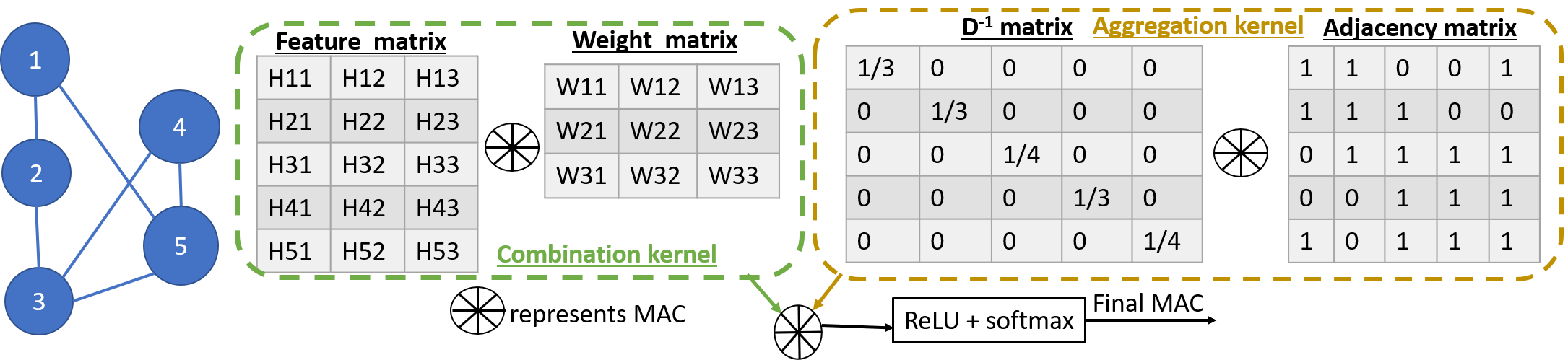}
\vspace{-2em}
\caption{\textbf{Undirected, unweighted graph with 5 nodes and 6 edges passing through 1-layer GCN. Combination showing MAC between dense feature and weight matrices, aggregation showing MAC between sparse D\textsuperscript{-1}, adjacency matrices to generate final MAC before ReLU, softmax function  }}
\label{fig:GCN_basic}
\vspace{-1em}
\end{figure*}
\section{BACKGROUND}
\subsection{Graph neural networks}
Graphs, characterized by nodes and edges, have been a traditional and crucial data structure to represent unstructured data across a diverse range of real-world applications. %The traditional graph processing (non-Machine learning) algorithms like travelling-salesman, max-cut problems traverse/partition the input graph nodes and solve a particular problem like identifying the minimum travel distance (travelling-salesman), maximum cut for partitioning (max-cut). 
%Recently, machine learning (ML) based graph processing has been researched to classify different nodes in a graph into pre-defined groups, clustering of graphical nodes, etc.
%The traditional deep learning algorithms like CNNs operate on structured data, and the irregular data patterns in graphs make it difficult to map them onto CNN for processing. 
In the pursuit of facilitating classification and clustering of graph nodes, considerable research effort has been devoted to Graph Neural Networks (GNNs). These networks are designed to extract distinctive features from graph structures and to enable an understanding of relationships among nodes within a graph. A variety of GNN variants, including  Graph Convolutional Networks (GCN), Graph Attention Networks (GAT), and GraphSage \cite{GAN}, \cite{GAN_1}, are being extensively researched. These explorations are geared towards unraveling specific attributes of interest in various domains such as social network modeling, molecular chemistry, citation networks, etc. %Different types of GNNs like \textbf{Graph convolutional networks (GCN)}, \textbf{Graph attention networks (GAT)} and \textbf{GraphSage} \cite{GAN}, \cite{GAN_1} have been researched to understand specific properties of interest in social network models, molecular chemistry, citation networks, etc. %Graph convolutional networks (GCN) based models have been of great interest, as they are CNN inspired models applied on graph dataset.  
%These models have been predominantly used to learn features by performing convolution/multiply and accumulate(MAC) followed .
A Graph Neural Network (GNN) model is composed of multiple layers. In each GNN layer, every node from the input graph undergoes processing by two kernels: the combination and the aggregation kernel.

 \par The function of the combination kernel is to convert the feature vectors (H), which characterize individual nodes in the graph, into an equivalent vector. This process entails the multiplication and accumulation (MAC) of the H values associated with each node in the graph using a weight matrix, akin to what is seen in a fully connected layer of a traditional neural network. The formulation of this kernel remains constant across all GNNs within the l\textsuperscript{th} layer and is expressed generically, for n\textsuperscript{th} node as:
%\begin{equation}
\textbf{H\textsubscript{comb}[l][n] = H\textsubscript{layer}[l-1][n]*W[l]}. When l=1, H\textsubscript{layer}[l-1] becomes same as input feature vector. For other layers, it is the final output of previous layer, similar to a traditional CNN. Across all nodes, feature vectors can be concatenated to form the feature matrix H\textsubscript{comb}[l] =\{H\textsubscript{comb}[l][n]\}
%\end{equation} 

\par The aggregation kernel combines attributes from neighboring nodes, represented by H, to gain insight into interactions with adjacent nodes and to mitigate data irregularities, by averaging across nodes \cite{GCN_GEOM}. This process varies across different GNN models. The aggregation mechanism for the l\textsuperscript{th} layer is denoted as \textbf{H\textsubscript{agg}[l] = M*H\textsubscript{comb}[l]}, with the matrix M being contingent upon the specific GNN model in use. The aggregation mechanisms associated with different GNN models are:
%This can be equivalently represented for n\textsuperscript{th} node in a layer as \textbf{h\textsubscript{agg-i}= $\Sigma$ C\textsuperscript{ij}*h\textsubscript{comb-i}}.
\par For \textbf{GCN}, M = D\textsuperscript{-1}*A, where D is the degree matrix normalized over node degree, typically represented as a diagonal matrix for easy matrix multiplication,  and A is the adjacency matrix (Fig.\ref{fig:GCN_basic}) across all neighbors. The adjacency matrix includes a self-loop, which results in the diagonal elements being set to 1. Similarly, the degree matrix is a diagonal matrix where D\textsubscript{ii} = $\Sigma$ A\textsubscript{i}.
\par 
Example of GCN: In Fig. \ref{fig:GCN_basic}, a depiction of a graph with 5 nodes in an unweighted, undirected setup is shown. Each node contains 3 features, resulting in a feature matrix size of 5*3. These features undergo a transformation using a weight matrix, akin to neural network convolutions. The weight matrix (combination kernel) is sized 3*3, leading to a resultant combination matrix of 5*3. Subsequently, aggregation follows, employing kernels consisting of the degree and adjacency matrix. For example, the feature vectors of node 1 (H11, H12, H13) transform based on the weight matrix. Node 1 has 3 adjacent nodes, accounting for self-looping. Consequently, D\textsuperscript{-1}\textsubscript{11} becomes 1/3. The aggregated output for node 1 is a scaled sum of combination vectors of nodes 1, 2, and 5.

\par In the case of \textbf{GraphSage}, M = D\textsubscript{samp}\textsuperscript{-1}*A\textsubscript{samp}, where D\textsubscript{samp} and A\textsubscript{samp} denote the sampled version of the degree and adjacency matrices respectively. In contrast to the approach of aggregating information from all neighbors, as seen in GCN, GraphSage exclusively employs a subset of chosen or pre-trained neighbors associated with a specific node for aggregation purposes. This selection process aims to ensure a consistent and predetermined count of neighbors for all nodes. Consequently, this strategy introduces a sense of relative preference among neighbors for aggregation.
\par For \textbf{GAT:}, M = (Attn)*(A), where Attn is the attention matrix. Unlike GCN and GraphSage, Attn relies on the combination vector of a node to determine the weight corresponding to each neighboring node. The Attn vector for a particular node is obtained by (i) Defining attention between i\textsuperscript{th} and j\textsuperscript{th} neighboring node as e\textsuperscript{a*(ReLU(H\textsubscript{comb-i}$||$H\textsubscript{comb-j}))} where a is a trained vector and $||$ indicates concatenation of combination vector of both the nodes (ii) Normalizing the attention across all the elements in a row of Attn matrix to refrain from numerical explosion.

%is a 2-step process with the adjacency matrix (A) multiplied with the resultant vector from combination phase to result in an intermediate vector, which is further scaled according to D\textsuperscript{-1} matrix (reciprocal of degree matrix) to obtain the resultant vector post aggregation, formulated as:

%\begin{equation}
 %  \textbf{H\textsubscript{l} = D\textsuperscript{-1}*A*H\textsubscript{comb-l}}
%\end{equation} 
ReLU is used as the activation function at the output of the aggregation layer to introduce non-linearity. Finally, softmax function is used for classification into different categories.

\subsection{Processing in/near-memory}
CPUs and GPUs have long served as the primary workhorses for various user applications, including GNNs. However, they face increased data movement costs due to their Von-Neumann architecture, requiring periodic data transfers between memory and computational units. Processing in Memory (PIM) aims to embed computations within memory, mitigating these costs. The choice of memory technology in PIM designs is crucial, with many GNN accelerators favoring ReRAM. However, ReRAM-based accelerators are often dedicated solely to GNNs, leading to increased dead silicon area when integrated into existing SOCs crowded with multiple accelerators. Understanding ReRAM characteristics is essential to grasp the drawbacks of these designs. 
\subsection{Resistive Random Access Memory (ReRAM) bitcell} 
This non-volatile memory element operates by varying resistance, unlike traditional charge-based memory like SRAM. It toggles between low-resistance (SET) and high-resistance (RESET) states, akin to '1' or '0' in SRAM/DRAM. The memory bitcell consists of 1T1R (1 Transistor 1 Resistor), wherein writing involves activating the corresponding word line (WL) and applying voltage to the bit line (BL) for SET/RESET. Reading relies on current flow, low for RESET and high for SET states. Despite their compactness, ReRAMs suffer from limited endurance, higher voltage requirements, latency, and susceptibility to process variations \cite{IGZOCIM} \cite{FeFET}\cite{UTBB_SOI}\cite{Cryo_arxiv}. Although, there are many drawbacks (which are overcome by SRAM bitcell), their compact nature is a driving factor \cite{PIM_GCN_2}\cite{NVM_Raman}\cite{JJFET}.
\subsection{Static Random Access Memory (SRAM) bitcell} 
SRAM stands out in cache/register file design for its ultra-low access latency (in the nanosecond range), surpassing other memory technologies \cite{Teja_SRAM}. Designs typically feature either a decoupled read/write port structure, seen in the 8-Transistor (8T) SRAM, or a shared read/write port structure, as seen in the 6-Transistor (6T) SRAM. The 8T SRAM employs WWL (Write word line) for writing and WBL (Write Bit Line) to store data in the storage node (S). During a read operation, RBL is precharged to a high voltage. If the S node holds '1', R2 is activated, discharging RBL through the R1-R2 stack. For a '0' in the bitcell, precharged RBL remains unchanged, aiding content distinction. The 8T SRAM performs efficient read-after-write (RAW) computations, unlike the typical 6T SRAM that necessitates write, precharge, and read commands, making RAW a 3-cycle operation. For 8T SRAMs, precharge is overlapped with the write command, thereby making RAW a 2-cycle operation, because RBL can be precharged, while a different row of bitcells is written using a combination of WWL/WBL. Also, these have lower write/read voltage/power as opposed to ReRAM, and resilience to process variations, thus giving performance, and power advantage \cite{ABI}\cite{NV_SRAM}. 
%This is a non-volatile memory element that leverages the principle of data storage based on resistance, unlike conventional charge-based memory elements like SRAM, etc. The device operates in 2 different resistance states - SET (low-resistance) and RESET (high-resistance), which essentially differentiates between contents stored in the bit-cell, equivalent to storing '1' or '0' in SRAM/DRAM \cite{NV_SRAM}. The easiest approach towards programming/writing memory bitcell, consisting of 1T1R (1 Transistor, 1 ReRAM), is by i) turning on WL (word line) corresponding to the row to be accessed in the memory array, and ii) driving BL (bit line) to positive/negative voltage for SET/RESET operation. To perform a read operation, BL is driven to a voltage that allows current flow through ReRAM. The current flow through ReRAM is small, when ReRAM is in RESET state, while it is high, when ReRAM is in SET state, thereby distinguishing 2 states. However, ReRAMs exhibit major drawbacks such as limited write endurance, elevated write voltage requirements, and increased write latencies \cite{Write_RRAM} \cite{IGZOCIM} \cite{FeFET} . Additionally, they are vulnerable to process variations, which in turn can lead to reliability issues. SRAMs overcome the major drawbacks and their bitcell characteristics are discussed next. Although these bitcells have multiple issues, the major motivating factor is the compact, dense nature of the bitcell/ReRAM based memory array \cite{PIM_GCN_2}. }
\begin{figure*}[t!]
\centering
\includegraphics[width=\linewidth]{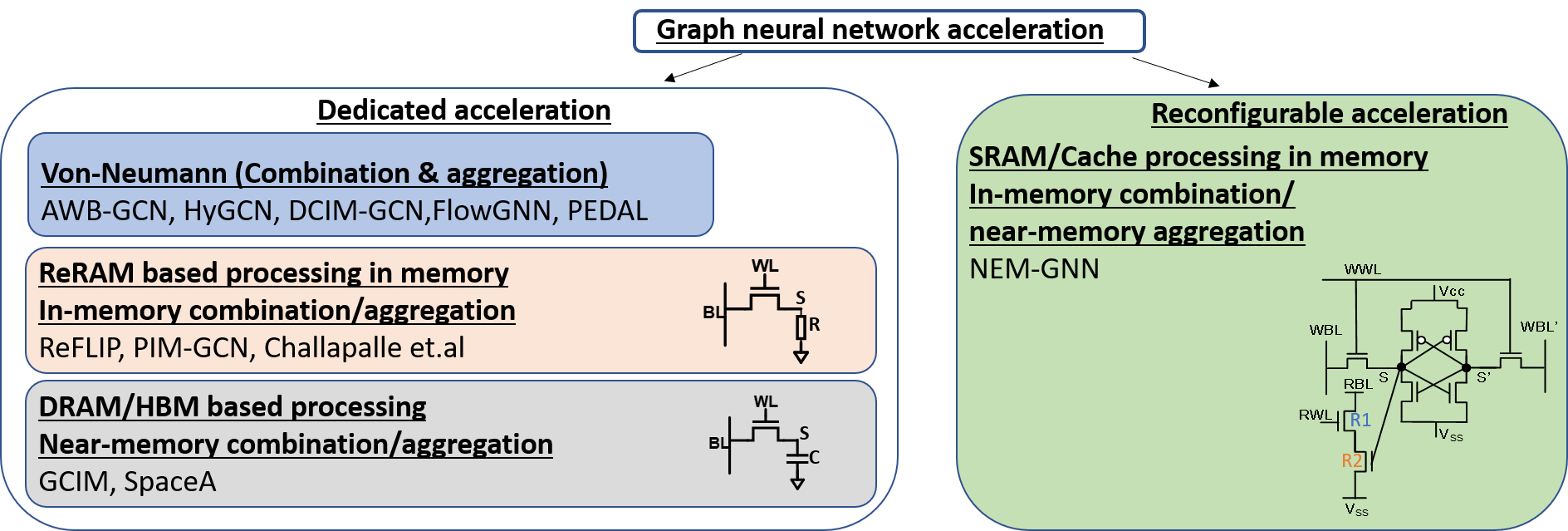}
\vspace{-2em}
\caption{\textbf{Landscape of Graph neural network based acceleration. The prior works are predominantly dedicated accelerators requiring periodic host-accelerator interaction. These are further classified into Von-Neumann, ReRAM based PIM, DRAM/HBM based PIM. The proposed accelerator is not dedicated and reuses cache in CPUs to perform GCNs. The bitcells for PIM designs are also shown }}
\label{fig:Motivation}
\vspace{-1em}
\end{figure*}

\subsection{Dynamic Random Access Memory (DRAM) bitcell}
DRAM serves as the main memory storage due to its low cost and compact bit-cell design. The 1T1C (1-Transistor, 1-Capacitor) bitcell (Fig.\ref{fig:Motivation}), utilizes a capacitor to store information. The write operation involves activating WL while storing data onto the 'S' node by driving BL. Precharging the BL to half of the operating voltage precedes the read operation, where the voltage at BL changes based on the bitcell contents. The advantages of DRAM lie in its compactness, cost-effectiveness, and high capacity. Additionally, its capacity can be augmented through 3D stacking, enabling high-bandwidth memory (HBM)/high memory cube (HMC) designs. DRAM faces challenges such as periodic refresh due to the capacitor's dynamic nature, resulting in performance/energy overhead.

\subsection{Related Work}
The existing GNN accelerators are classified into Von-Neumann/PIM-based dedicated accelerators. Von-Neumann-based dedicated architectures like AWB-GCN\cite{AWB-GCN}, HyGCN \cite{Hygcn} and DCIM-GCN\cite{DCIM_GCN}, PEDAL \cite{PEDAL}, FlowGNN \cite{FlowGNN}  make use of dedicated accelerator, specifically designed for GNN. The logic units are present outside the memory for performing combination and aggregation. %It's important to note that these accelerators are specifically optimized for GCN and aren't designed for other GNNs.
\par \textbf{AWB-GCN} introduces hardware-level enhancements aimed at addressing the challenge of workload imbalance, a consequence of the interaction between sparse input graphs and dense weight matrices. The proposed hybrid architecture employs distinct storage units for the combination and aggregation stages. This is complemented by specialized control logic that facilitates the seamless distribution of workloads across different processing engines. However, it's important to note that this strategy brings about an additional area overhead and offers only a modest assurance of optimal resource utilization. Moreover, the scalability of this solution is constrained by the number of processing elements. This is geared towards optimizing GCN predominantly.  \par \textbf{HyGCN} introduces an architectural approach and programming paradigm that harnesses both intra-vertex and inter-vertex parallelism during the combination and aggregation phases, thereby enhancing overall performance. Nevertheless, it's worth noting that despite these improvements, HyGCN lacks an inherent awareness of sparsity, leading to less efficient utilization of hardware resources and consequently resulting in an additional area overhead. Moreover, the inherent limitations of the Von-Neumann architecture become evident as graph sizes grow, contributing to the energy overhead caused by frequent data transfers between memory and computational units. \par \textbf{DCIM-GCN} \cite{DCIM_GCN} employs memory as a storage component and supplements it with NOR gates/logic in proximity to the memory for near-memory combination. The aggregation, on the other hand, utilizes the Von-Neumann architecture and is specifically optimized for GCN.  
\par \textbf{PEDAL} \cite{PEDAL} introduces a power-efficient dataflow accelerator that optimizes dataflow based on the incoming graph's nature, enhancing efficiency and flexibility, by changing the order of execution between combination and aggregation. It employs a dedicated Von-Neumann architecture, necessitating periodic CPU-accelerator interaction.

\par \textbf{FlowGNN} \cite{FlowGNN} proposes a generic accelerator with parallelism ranging from node/edge, using multiple data queues, execution units along with scatter/gather mechanisms. The dataflow is general and can be applied to any graph based model like GCN, GAT, GraphSage. This architecture is realized on a FPGA, similar to other Von-Neumann accelerators, and suffer from similar drawbacks
%For instance, in Citeseer, with 3,700 features and 3,300 nodes with 1-layer GCN with weight matrix of size 3700*10, capable of classifying the input graph nodes across 10 different types, the number of MACs is approximately equal to 3700*10*3,300 nodes. The overall energy can be broken down into energy for data movement (including the access energy for memory), energy for computation. The access energy assuming SRAM energy of $\sim$5fJ/bit, the access energy would amount to 10\textsuperscript{3} times the energy access of a single bit, for a single bit weight, assuming the SRAM is accessed only once for performing all MACs. The number of graph nodes, H size and access energy are the motivators behind performing PIM.
\par PIM-based accelerators predominantly have utilized ReRAM/DRAM for in-memory processing. ReFLIP, PIM-GCN \cite{PIM_GCN}, Chappalle et.al \cite{PIM_GCN_2} present Processing-in-Memory (PIM) accelerator that utilizes a crossbar ReRAM architecture. \textbf{ReFLIP} adopts a weight stationary approach for executing dot product operations, and it includes a peripheral DAC/ADC to perform analog compute. However, there are notable drawbacks associated with this approach, which are detailed in Sec.3. \par \textbf{PIM-GCN} uses a similar approach to ReFLIP for performing MAC for combination, except that the aggregation is performed using a 2-step process with the first step being a content-address-memory (CAM) to identify neighbors, and the second step involves performing MAC. The proposed approach aims to schedule node computations in a way that the inter-node parallelism is maximal.
\par \textbf{Challapalle et.al} propose multiple engines with an architecture similar to PIM-GCN, except that the presence of separate engines potentially aids performance, at the cost of power/energy. However, the architecture is limited to performing GCN, cannot be extended to GAT/GraphSage
\par \textbf{DRAM/HBM} based near-memory accelerators make use of compute units closer to DRAM. \textbf{GCIM} \cite{GCIM} uses a novel data-aware mapping algorithm to efficiently utilize near-memory (3D-stacked HMC) MAC units. \textbf{SpaceA} \cite{Spacea} uses a near-memory CAM/MAC structure in processing engines to enable graph processing and perform workload balancing by mapping different sparse features to different banks, and is optimized for general graph processing and not for GCNs. 
\section{MOTIVATION}
In this section, we motivate NEM-GNN by highlighting the issues in the existing PIM designs. The combination phase, predominantly consisting of convolution between weights and feature vectors, is naturally suitable for PIM compute. Furthermore, the dense weights are reused across various feature vectors, establishing the PIM-based combination as a weight-stationary approach. 
Aggregation is not amenable to PIM compute, as detailed later.

\begin{figure*}[t!]
\centering
\includegraphics[width=\linewidth]{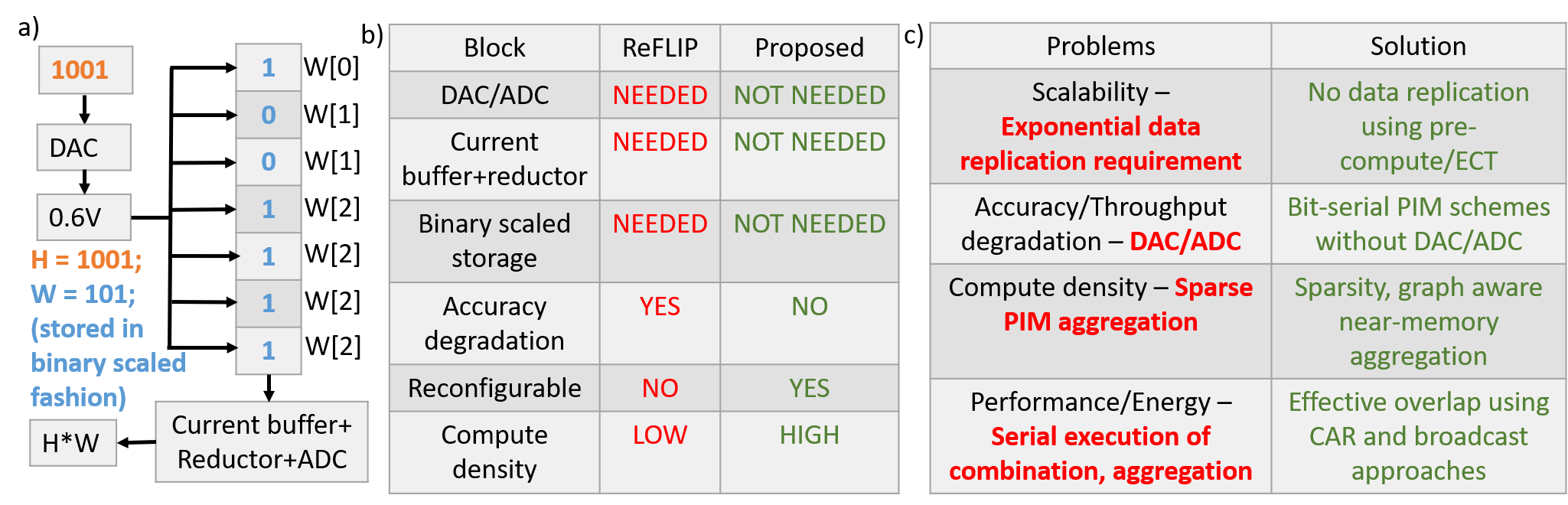}
\vspace{-2em}
\caption{\textbf{a) ReRAM approaches (i) use DAC for incoming H conversion to an equivalent analog value (ii) store weights of GNN in binary scaled fashion (iii) utilize current buffer+reductor to perform current-based summation and ADC to generate H*W  b) Qualitative comparison between ReRAM approaches and NEM-GNN c) A summary of the identified issues and the proposed solutions}}
\label{GCN_SRAM}
\vspace{-1em}
\end{figure*}
\subsection{Issues in ReRAM-based PIM for combination}

\par ReRAM device limitations are discussed in Sec.2.3 and also detailed in \cite{RRAM_cache}. This section covers disadvantages, with specific reference to compute, as the existing ReRAM based accelerators follow the same strategy for performing in-memory dot-product compute, as that of ReFLIP. Fig.\ref{GCN_SRAM}a) illustrates dot product compute between H=1001 and W=101. The weight bits are stored in a binary scaled fashion: the 0\textsuperscript{th} bit is stored once, the 1\textsuperscript{st} bit is replicated twice, and the 2\textsuperscript{nd} bit is replicated four times. In the combination phase, the corresponding analog value for H is mapped onto the word line (WL), and the current flowing through the bit-line is aggregated using a current reductor. This is then fed into an Analog-to-Digital Converter (ADC) to derive the equivalent digital value.  Firstly, the weights are stored in binary scaled fashion in memory. The incoming H value of 1001 is converted into an equivalent analog voltage of 0.6V by using DAC. This voltage then activates multiple bit-cells in the same column, by mapping H onto WL. The current summation across all ReRAM bitcells in a column (BL) is then performed by a combination of the current buffer and reductor. This is then converted into an equivalent voltage, and fed into ADC, which outputs an equivalent digital value of 101101. NEM-GNN overcomes the usage of analog blocks (summarized in Fig.\ref{GCN_SRAM}b). The disadvantages of the compute strategy are: (i) The presence of ADC and using ReRAM-based compute implies that the existing SOC components cannot be reconfigured to realize ReFLIP and need a dedicated accelerator arrangement.
(ii) The binary scaled storage requirement requires that 2\textsuperscript{n}-1 rows are needed for storing an n-bit weight. This implies that the area of the memory array scales exponentially as the precision of weight increases, limiting scalability.
(iii) ADC should be extremely precise, as an 8-bit weight multiplied by an 8-bit feature vector (H) requires a 16-bit ADC. The process variations inherent to ADCs pose serious implications on the compute accuracy (Fig.\ref{GCN_SRAM}b).
(iv) Throughput is restricted by the number of ADCs per bank (1 in the case of ReFLIP). Furthermore, the ADCs and current reductor affect the energy of the design with an additional area overhead arising from bulky ADCs. 
(v) Compute density, quantified as the ratio of operations performed within memory per bit-cell, serves as an indicator for gauging the efficiency of hardware resource utilization. In ReFLIP, for the dot product between an m-bit feature vector and an n-bit weight, the compute density equals (m*n)/(2\textsuperscript{n}), which needs to be improved substantially. These issues are summarized in Fig.\ref{GCN_SRAM}c)

\subsection{Issues in ReRAM-based PIM for aggregation}
ReRAM-based PIM approaches perform both combination and aggregation in memory. However, aggregation is not really suitable for PIM computation, because of the following reasons: Firstly, realizing the exponential compute required for aggregation, like in GATs, is not suitable for PIM. \\ Secondly,  aggregation in memory can be achieved by 2 means: 
 (i) The first option involves the usage of a separate memory array to store the resultant dot product from the combination phase (ii) The second option involves the reuse of the existing combination memory array for aggregation as well. The first option incurs additional area overhead for the associated memory array resulting in ineffective utilization of hardware resources, degrading the array density, and energy for compute. The second option improves the utilization of hardware resources at the cost of performance and additional control circuitry. Furthermore, this adds a degree of serialization between combination and aggregation, as aggregation can be initiated only after combination is complete. This causes a performance bottleneck with no overlap between combination and aggregation. Therefore both these approaches are inefficient in terms of performance, energy, and area. ReFLIP and PIM-GCN uses the latter, while Challapalle et.al use the former. Finally, the sparse aggregation compute, if performed in-memory degrades the compute density, as most PIM computes are insignificant.
\subsection{Issues in DRAM-based near-memory compute for GNN}
DRAMs, designed for cost efficiency and expanded storage, face constraints when integrating near-memory processing, leading to reduced storage capacity \cite{ISCAS_IGZO}\cite{DRAM_arxiv}\cite{UT_Thesis}. Given that DRAM is a single-chip package from manufacturers and directly integrated into existing SoC architectures, the fixed area allowance must accommodate additional near-DRAM logic. Notably, Samsung's near-DRAM chip for ML acceleration halved its storage capacity to include per-bank near-memory logic \cite{Samsung}. Consequently, traditional DRAM-hits turn into (DRAM+PIM)-misses, causing performance dips in storage-intensive workloads. This issue persists even in HBMs, with smaller storage capacities (e.g., recent HBM3 - 24GB vs. DDR4 - 256GB) and reduced storage density due to added bulky near-memory digital logic to DRAM bitcell arrays.
\par From GNNs' perspective, large datasets can push DRAM capacities to their limits. Storing basic node information (excluding edges/weights) with an 8-bit representation for features in PubMed/Reddit datasets alone demands 0.4GB/0.5GB. Preserving DRAM capacity becomes crucial to avoid the consequential power/energy expenses. Reduced DRAM capacity might limit concurrent CPU applications, and performance degradation in traditional CPU workload execution. 
\par Samsung's demonstration focused on ML, optimizing for MAC operations. However, GNNs differ as aggregation navigates nodes, utilizing CAM operations, distinct from mere arithmetic operations. There are two approaches: i) Employ CAMs via traditional CPU, causing GNN execution degradation due to CPU-DRAM interaction, different from ML applications. ii) Including extra CAMs would further reduce storage capacity. Considering XNOR computation area vs. FP16 addition, this logic could trim DRAM by 10\%. For instance, a reduced 3GB HBM-based PIM (originally 6GB, halved for ML in \cite{Samsung}) now reduced to 2.7GB, pushes DRAM limits, potentially causing misses, impacting performance/energy for datasets. Moreover, PIM's 3.5x ML performance gain would see a 10\% reduction. Moreover, SpaceA augments HBM with specialized processing engines, expanding bit-width, incorporating CAMs, and load queues. This adaptation reduces DRAM capacity, impacting both traditional CPU tasks and GNN performance. GCIM adopts 3D-stacked HMC, encountering issues akin to SpaceA, facing bandwidth constraints leading to Micron halting its production in 2018. To summarize, Compute-near-DRAM approaches reduce DRAM capacity, causing performance and energy bottlenecks with increased misses, particularly for high-capacity workloads.
\begin{figure}[t]
\centering
\includegraphics[width=12cm]{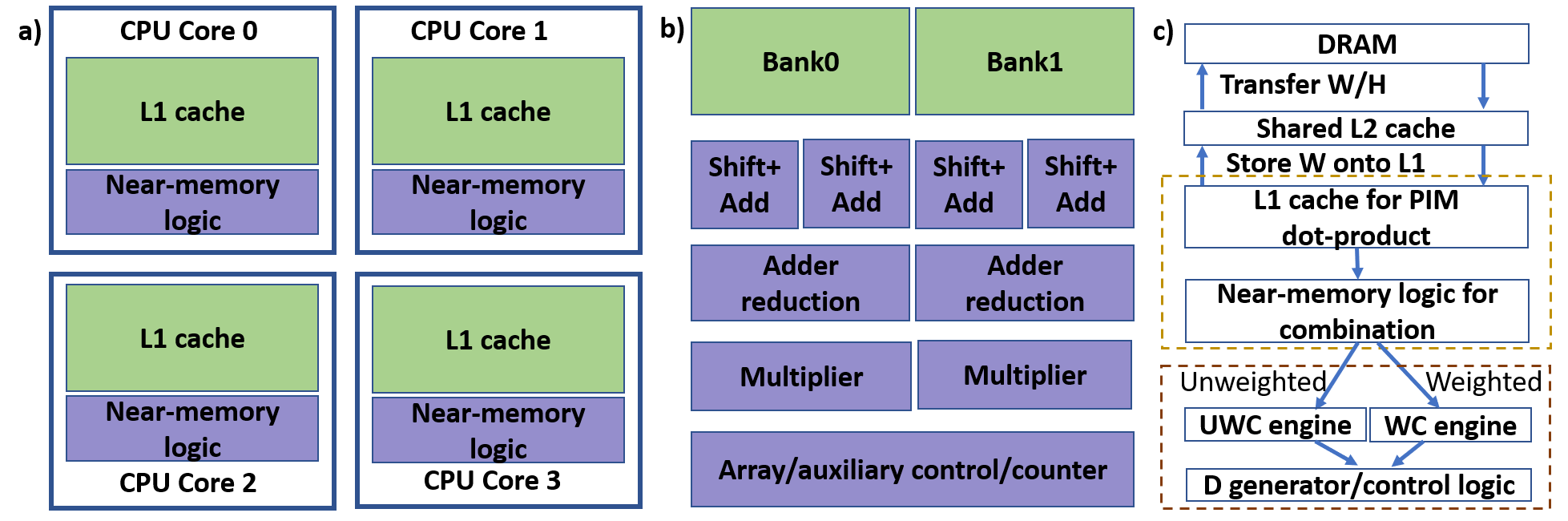}
\vspace{-1em}
\caption{\textbf{a) NEM-GNN is realized by repurposing the L1 cache for in-memory compute, with minimal near-memory peripheral logic added to each CPU core. b) In an L1 cache, consisting of 2 banks, shift and add are present at a granularity of 1 per every 8 columns per bank, with 1 adder reduction/multiplier per bank, and other dedicated logic shared across the entire cache. c) DRAM is accessed to transfer weights/ feature vectors into shared L2 cache. L1 cache stores weights. Combination datapath involves PIM (L1 cache) dot-product with near-memory logic. The combination result goes via the UWC (Unweighted control)/WC (Weighted control)  engine depending on the weighted nature of the underlying graph, followed by D-generator/control logic for aggregation. It is to be noted that UWC/WC engines use the added near-memory logic (multiplier/adder) and do not require dedicated hardware. }}
\label{Overall_organization_1}
\vspace{-1em}
\end{figure}
\section{NEM-GNN's salient features}
%In this section, we give an overview of our approach to addressing the concerns, listed in Section 3. 
\subsection{Reconfigurability}
The requirement of dedicated accelerator arrangement in previous works is overcome by reusing the L1 cache inside the CPU core for performing in-memory compute with additional near-memory logic. Fig.\ref{Overall_organization_1}a) shows a multi-core CPU design with each core integrating dedicated minimal near-memory logic for performing combination/aggregation. Fig.\ref{Overall_organization_1}b) shows the per-core additional near-memory logic for L1 cache, assumed to be consisting of 2 banks for illustration. Shift and add are present at a granularity of 1 for every 8 columns (assuming weights for 8 bits) for every bank. 1 adder reduction tree/multiplier per bank along with buffer/control/counter shared across the entire L1 cache is used to realize NEM-GNN.  
\par The datapath is split into combination and aggregation. For the combination data path, the compute array reuses the L1 cache to achieve PIM functionality with additional near-memory control logic for early compute termination and addition. For aggregation, we utilize a near-memory buffer for storing the adjacency matrix, a D generator for generating the corresponding degree(M) matrix, and UWC/WC engines for graph/sparsity-aware aggregation. If the underlying graph is unweighted, the UWC engine is used for aggregation, while the WC engine is used for weighted graphs. It is to be noted that these engines reuse the added near-memory logic like multiplier/adder and do not require dedicated logic for their functionality. There is no requirement for ADC/DAC/specific memory technology, enabling integration of NEM-GNN into a traditional CPU pipeline. For GNNs that do not fit on-chip, DRAM is accessed to fetch data into L1/L2. The access latency of DRAM is amortized by prefetching data, as PIM accesses are deterministic. 
 \subsection{Scalable L1 cache digital PIM compute without DAC/ADC - Architecture to circuit}
 Scalability is affected by exponential area requirement as the precision of weight increases. This is primarily because of the requirement for data replication across different banks. We propose 3 PIM approaches, which begin with n\textsuperscript{2} data replication (NEM-C1), and further reduce it to no replication requirement (NEM-C2/C3), as detailed in the next section. 
 \par
\begin{figure}[t]
\centering
\includegraphics[width=12cm]{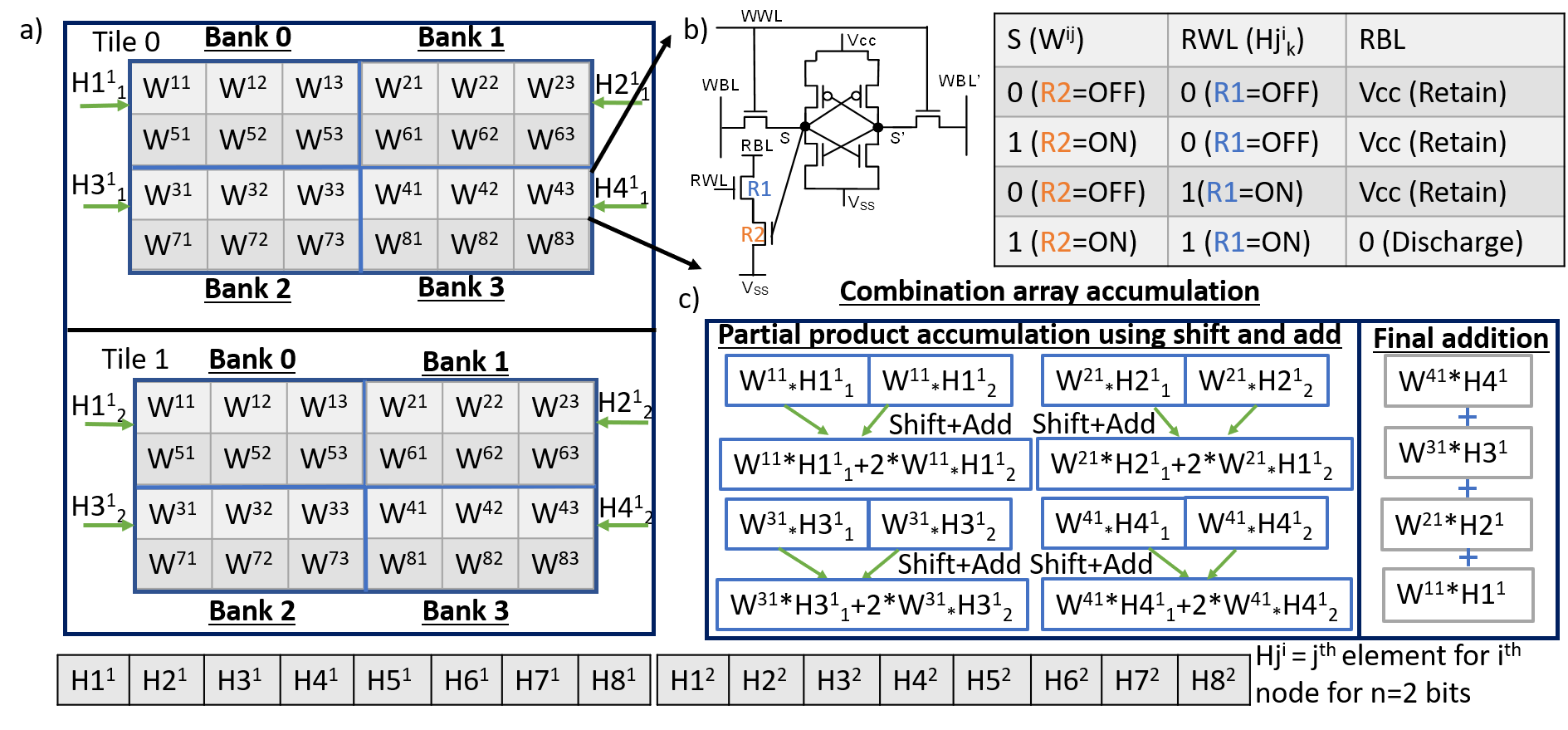}
\vspace{-1em}
\caption{\textbf{ a) Compute array organization for NEM-C1: 2 tiles with 4 banks in each tile, with bit-serial PIM  performed between H mapped onto RWL and W replicated across both tiles is shown for illustration. 2-bit 8-element H and 1-bit 8*3 weight matrix is shown with Hj\textsuperscript{i}\textsubscript{n} indicating n\textsuperscript{th} bit of j\textsuperscript{th} element for i\textsuperscript{th} node. b) W is stored in 8T SRAM bitcell in L1 cache, and H is mapped onto RWL. RBL discharge is used as a measure of dot product between W and H. RBL is initially precharged to Vcc and discharges only when W and H are '1'. c) Shift and add across partial dot products from PIM compute, with the result of final addition written into a row of combination array}}
\label{Overall_organization}
\vspace{-1em}
\end{figure}

 Our proposed designs implement a digital bit-serial PIM methodology for combination that eliminates the necessity for DAC/ADC. This setup proves robust against process variations, ensuring accurate computations. With the employment of bit-serial computation devoid of DAC/ADC, the design's throughput is freed from ADC-related constraints. The bit-serial computation, in conjunction with strategies like early computation termination or pre-computation, facilitates achieving high performance/throughput for NEM-GNN designs. Moreover, the absence of data replication combined with bit-serial computation contributes to an enhanced compute density, elaborated in Section 5. %These designs use a bit-serial PIM approach without any requirement of DAC/ADC, thereby being resilient to process variations, resulting in accurate compute. The bit-serial compute without DAC/ADC implies that the throughput of the design is not ADC bound. The bit-serial compute along with strategies such as early compute termination/pre-compute helps achieve high throughput in NEM-GNN. Furthermore, the requirements of no data replication along with bit-serial compute helps achieve improved compute density, as discussed in the next section   
  %\par \textbf{\underline{Reconfigurability}} is important, as it enables accelerating GNNs with decreased overhead when GNN-specific hardware is added onto the existing SOC designs. AWB-GCN, HyGCN are envisioned as dedicated accelerators connected externally to a CPU, requiring frequent host-accelerator interaction and data movement costs. DCIM-GCN requires additional NOR gates in every column of the memory array which are not present in the existing SOC/CPU caches/memory arrays. ReFLIP requires additional DAC/DAC on the periphery of the memory array in the ReRAM-based in-memory design (Fig.\ref{fig:Motivation}a). Therefore, the existing designs have additional costs, when compared to NEM-GNN, which can be reconfigured from SOC components. 
  %\subsection{L1-cache PIM for combination} 
  Furthermore, given that i) ReRAMs suffer from scalability, compute density, performance/energy issues, ii) DRAMs suffer from reduced storage capacity, we propose L1-cache based PIM, that retains storage of DRAM, while offering improved performance/energy.  From a memory organization standpoint, L1 cache consists of multiple tiles, with each tile consisting of multiple banks, leveraging tile and bank-level parallelism (as shown in Fig.\ref{Overall_organization}a). These designs use decoupled read/write port structure (8T SRAM), with the decoupled read port transistors marked as R1, R2 in Fig.\ref{Overall_organization}b). The read port transistors are re-purposed to perform dot product compute (bit-wise AND) for combination by mapping H bits onto RWL and storing weights in the bitcell. The compute can be described as follows: (i) RBL is precharged to Vcc before compute. (ii) Only when both S and RWL are '1', RBL discharges via the read-port transistors (R1/R2) implying that the computed value is 1. (iii) RBL remains at Vcc for other combinations of H and W, because one of the read-port transistors are turned OFF. When W='0', R2 is turned OFF, as the S node is 0; When H='0', R1 is turned OFF, preventing discharge of RBL in either case. Using this PIM approach, we propose 3 different bit-serial scalable, performance and compute density optimized PIM approaches namely (a) n\textsuperscript{2} data replication, (b) no data replication with early compute termination (ECT), (c) pre-compute, which are detailed in NEM-C* section.
  
  %which cannot be achieved by reconfiguring the existing SOC memory arrays.%For instance, in Citeseer, with 3,700 features and 3,300 nodes with 1-layer GCN with weight matrix of size 3700*10, capable of classifying the input graph nodes across 10 different types, the number of MACs is approximately equal to 3700*10*3,300 nodes. The overall energy can be broken down into energy for data movement (including the access energy for memory), energy for computation. The access energy assuming SRAM energy of $\sim$5fJ/bit, the access energy would amount to 10\textsuperscript{3} times the energy access of a single bit, for a single bit weight, assuming the SRAM is accessed only once for performing all MACs. The number of graph nodes, H size and access energy are the motivators behind performing PIM.  

\subsection{\textbf{Graph and sparsity-aware aggregation}}
 We use a near-memory approach for aggregation, that enables performing exponential in-memory compute. The near-memory aggregation with CAR and broadcast approach helps achieve better performance in NEM-GNN, as aggregation latency is effectively hidden by computing as soon as the combination results are ready. This is done by effective broadcast of combination results, as soon as the combination outputs are available for aggregation. 
 \par Furthermore, the sparsity-aware approach helps alleviate insignificant computations, while leveraging PIM's high throughput for performing significant computations. Graph connectivity-aware aggregation helps optimize aggregation based on different graphs, enabling achieving high performance, without area overhead, for graphs that fundamentally require less computation for aggregation. For instance, unweighted graphs require less computation as compared to weighted graphs, because of no requirement on multiplying with weight of the graph edge for aggregation.     % Fig.\ref{fig:Motivation}c) shows that the aggregation latency in NEM-GNN is 30-45x times that of ReFLIP. %  Moreover, performing aggregation in memory would imply 2 possible scenarios: (i) Usage of a separate memory array to store the resultant dot product from combination (ii) Reuse the existing combination compute array for aggregation. (i) incurs additional area overhead for the memory array and associated peripheral logic like DAC, ADC, with ineffective utilization of hardware resources, degrading the array density, energy (ii) Improves the utilization of hardware resources at the cost of performance and additional control circuitry, making both approaches PEA in-efficient. %In this article, we propose a near memory architecture that is area/energy/performance efficient for aggregation, by re-purposing hardware resources and overlapping computations with combination phase.

\subsection{No impact on traditional CPU workloads}
Since, NEM-GNN is realized reusing L1 cache, with additional near-memory logic, with no reduction in storage capacity, there is no impact on traditional CPU workloads. Firstly, there are no SRAM memory array level modifications to realize NEM-GNN, that could potentially impact the execution of traditional CPU workloads. Secondly, there is minimal near-memory logic that is added to realize NEM-GNN, achieved through effective reuse of hardware. For instance, the UWC/WC engines share the same adder, D-generator shares the multiplier with UWC/WC engines, and softmax shares exponential compute, with aggregation for GAT. Thirdly, the additional near-memory logic does not intervene in normal CPU operation and does not add additional power overhead. 

\section{NEM-C*: SCALABLE IN-MEMORY COMBINATION DESIGNS WITHOUT DAC/ADC}
\subsection{NEM-C1: Scalable bit-serial PIM with n\textsuperscript{2} data replication} 
\par The exponential storage requirement (n-bit weight requiring 2\textsuperscript{n} bitcells, independent of H bit-width) in ReFLIP is combatted by bit-serial PIM, with n-bit weight requiring m*n bitcells for m-bit H, as detailed in the subsequent paragraph. 
\par %Fig.\ref{Compute_tile_org} shows the convention used for a 2-bit, 8-element feature vector and 1-bit 8*3 weight matrix (W). 
 At the bank/tile level, the weights are kept stationary in the compute arrays, making the overall design a weight-stationary design. The weights are shown to be replicated for illustration of NEM-C1 design in Fig.\ref{Overall_organization}a) (while other designs do not require data replication). 8-element H vector each of 2-bits and 1-bit 8*3 weight matrix is shown with Hj\textsuperscript{i}\textsubscript{n} indicating n\textsuperscript{th} bit of j\textsuperscript{th} element for i\textsuperscript{th} node. The observation is that during dot product compute between H for each node (feature vector) and weight matrix to output combination vector, the same H element (H1\textsuperscript{1}) is used across a row of weights (W\textsuperscript{11}, W\textsuperscript{12}, W\textsuperscript{13}). This re-use is used to map a row in the weight matrix onto a row in a compute bank, with H mapped onto the corresponding RWLs in a bit-serial fashion, enabling W\textsuperscript{xy} parallelism over y-dimension (H1\textsuperscript{1}\textsubscript{1}* W\textsuperscript{11} computed in parallel with H1\textsuperscript{1}\textsubscript{1}* W\textsuperscript{12}). Inside a bank, eg. in bank0, H1\textsuperscript{1}\textsubscript{1} and H5\textsuperscript{1}\textsubscript{1} are mapped in successive cycles onto the RWL as a row is computed per bank, per cycle. The weight matrix is split across banks in a tile, enabling Hj\textsuperscript{i}\textsubscript{n} parallelism over j-dimension (H1\textsuperscript{1}\textsubscript{1}*W\textsuperscript{11} computed in parallel with H2\textsuperscript{1}\textsubscript{1}*W\textsuperscript{21}) and the weights are replicated across tiles for enabling Hj\textsuperscript{i}\textsubscript{n} parallelism over n-dimension (H1\textsuperscript{1}\textsubscript{1}*W\textsuperscript{11} computed in parallel with H1\textsuperscript{1}\textsubscript{2}*W\textsuperscript{11}). The replication of weights across multiple tiles ensures that the dot product between Hj\textsuperscript{i}\textsubscript{n} and W\textsuperscript{xy} is computed using O(n\textsuperscript{2}) bitcells in a single cycle. 

 \par This is illustrated by using an example in Fig.\ref{Overall_organization}. If the different bits of Hj\textsuperscript{i} are mapped onto RWLs of the same banks in different tiles, then bank0 of tile0 computes W\textsuperscript{11}*H1\textsuperscript{1}\textsubscript{1}, while bank0 of tile1 computes W\textsuperscript{11}*H1\textsuperscript{1}\textsubscript{2}. These 2 partial dot products can be shifted and added together to generate the dot product W\textsuperscript{11}*H1\textsuperscript{1}. Similarly, dot products obtained across different banks and tiles are shifted and added to obtain W\textsuperscript{21}*H2\textsuperscript{1}, W\textsuperscript{31}*H3\textsuperscript{1}, and W\textsuperscript{41}*H4\textsuperscript{1}. Finally, a full addition is performed to generate the MAC result for combination.
 %Near-memory shifter and adder (SA) gives the dot product result between an m-bit H and n-bit weight. %Fig.\ref{Overall_organization}b) assumes 1 SA per bank, enabling parallel dot product compute of 4 weights (W\textsuperscript{(1-4)}1) and 4 H elements (H(1-4)\textsuperscript{1}). If a row of SAs are used per bank, then dot between W\textsuperscript{(1-4)}(1-4) and (H(1-4)\textsuperscript{1-4}) is computed in parallel. 
 Full adder accumulates dot product from SA into a row of combination array (Fig.\ref{Overall_organization}c). 
\par The overall replication requirement is reduced to m*n for dot product between m-bit H and n-bit W, resulting in O(n\textsuperscript{2}) data replication. Furthermore, the compute density for the L1-cache compute is improved from (m*n)/(2\textsuperscript{n}) in ReFLIP to ((m*n)/(m*n)=1) in NEM-C1, indicating effective utilization of hardware resources. Moreover, the lower overhead of SAs compared to DAC/ADC, and the lesser number of RWLs to be turned ON (n\textsuperscript{2} vs 2\textsuperscript{n}) leads to reduced energy requirement.

\begin{figure*}[t!]
\centering
\includegraphics[width=\linewidth]{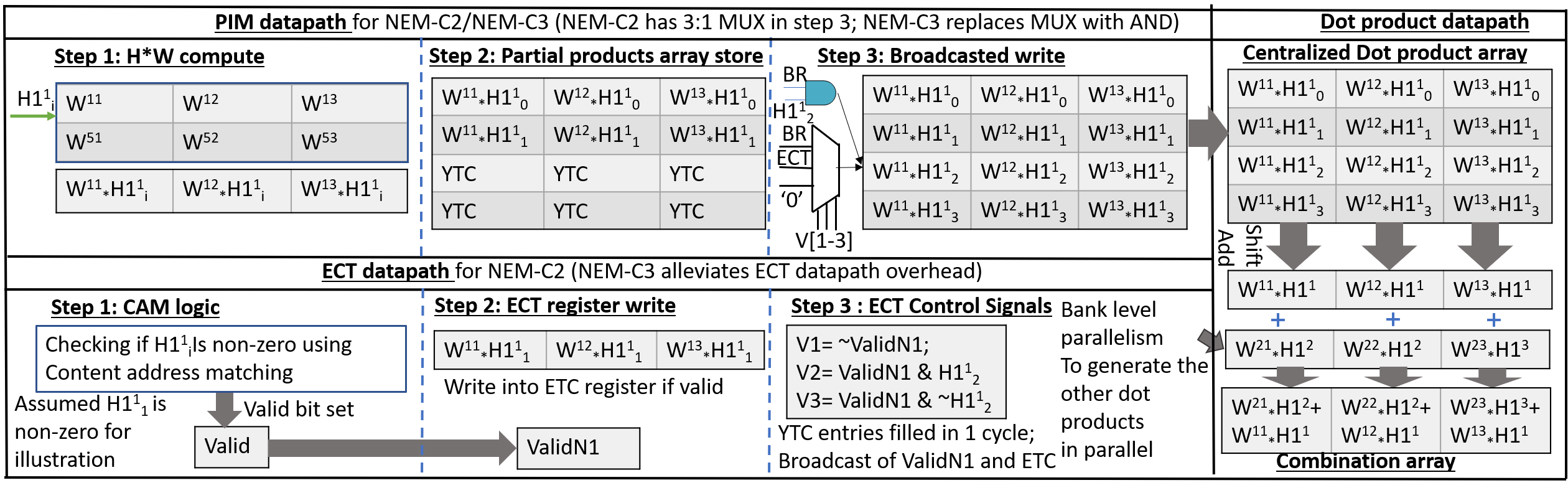}
\vspace{-2em}
\caption{\textbf{\underline{NEM-C2}: Early compute termination (ECT) occurs once one of the bit-serial H element bits is found to be 1, without data replication requirement. ECT data path checks for non-zero H bit in step 1 and writes the non-zero dot product into ECT register in step 2. In parallel, PIM datapath computes partial dot products in step 1 and subsequently stores them in the ECT register in step 2. This value is broadcasted to fill the other dot product entries in step 3 of PIM datapath using ECT control signals generated in step 3 of ECT datapath. % Bit serial PIM approach with early compute termination. PIM and ECT data path are executed in parallel to generate the final dot product. The dotted vertical lines indicate the different stages to generate the dot product for a 4-bit feature vector with n-bit weight (as n-bits are executed in parallel). 
3:1 MUX for writing into the partial products array: BR/ECT/'0' indicate bank read (computed value from memory)/ read of ECT register/ zero value, respectively.  Dot product datapath fills the combination array with the result of combination. \underline{NEM-C3} eliminates ECT data path by pre-computing for H bit being 0/1, uses AND gate (marked in blue) instead of 3:1 MUX in step 3, while using PIM and dot product datapaths, similar to NEM-C2}}
\label{fig:ETC_CIM}
\vspace{-1em}
\end{figure*}
\subsection{NEM-C2: Scalable bit-serial PIM with Early Compute Termination (ECT)}
\par NEM-C2 aims to eliminate the need for weight replication in the compute array by utilizing previously computed values. This aims to enhance the efficiency of bit-serial computation in terms of area, energy, and compute density. For example, when storing 3700*10 weights for a single layer of GCN in memory, where both feature vectors and weights are represented with 8-bit resolution, ReFLIP would demand about 1.2MB of storage, NEM-C1 would need around 0.2MB, and NEM-C2 would utilize only 4.6KB for storing these weights.
\par Similar to NEM-C1, multi-bit weights are stored in a single row of compute array, with elements of H mapped onto RWL in a bit-serial fashion, enabling  W\textsuperscript{xy} parallelism over y-dimension. Various banks serve to accommodate distinct weight rows, thereby permitting Hj\textsuperscript{i}\textsubscript{n} parallelism across the j-dimension. Furthermore, parallelism across the i-dimension extends across tiles. There is no requirement for data replication across banks, unlike NEM-C1. \par Fig.\ref{fig:ETC_CIM} illustrates the interplay among distinct data paths: PIM, ECT, and the dot product path. In step 1 of PIM datapath, the "H*W compute" stage generates the dot product between a row of weights and a bit of H element. Subsequently, in the second step of the PIM data path, the resulting dot product between H1\textsuperscript{1}\textsubscript{i} and (W\textsuperscript{11}, W\textsuperscript{12}, W\textsuperscript{13}) is stored in the partial products array. It's notable that when one of the bits within the bit-serial H element is determined to be 1, it leads to the identification of both potential outcomes from multiplying a row by 0 and 1 (as multiplying by 0 would output 0). This knowledge can be leveraged to prematurely terminate the H*W computation, thus saving energy. The ECT datapath provides assistance to terminate compute early, by identifying the position of '1' in a bit-stream of H mapped onto RWL. The Content-Address Matching (CAM) logic, allows for rapid searching based on content rather than memory addresses. 
%It traditionally stores both data and its associated addresses simultaneously, enabling quick retrieval based on the stored content. This feature is beneficial in applications requiring high-speed search operations.} 
This is utilized in step 1 of ECT data path in Fig.\ref{fig:ETC_CIM} is used to identify whether H-bit is '1'. A valid bit is set if a non-zero element is found. The associated dot products between the non-zero H1\textsuperscript{1}\textsubscript{1} and (W\textsuperscript{11}, W\textsuperscript{12}, W\textsuperscript{13}) are written into the ECT register in step 2, and the partial products array in parallel. Upon the completion of the ECT write operation, the remaining elements of the partial products array (denoted as YTC - Yet to Compute) are promptly filled by employing a single-cycle, broadcasted write using the data contained within the ECT register. This broadcast begins with the generation of ECT control signals in step 3 of the ECT data path, using the values of H elements and valid bit indication from CAM logic. Specifically, "BR" is chosen if the valid bit is '0', "ECT" is chosen if both the valid bit and the corresponding H bit are '1', and '0' is chosen if the valid bit is set while the corresponding H bit is '0'. The broadcast is done in step 3 of PIM datapath, using a 3 write-ported partial products array with 3 ports being bank read (when CAM logic has not detected a 1 in H element yet), ECT read (when ECT register is filled), and '0' (when H element bit is 0).  %ECT relies on the fact that all bits of H are readily available from the read-out of the storage array for control signal generation. 
Finally, the partial dot products are accumulated from the dot product array, onto the combination array, which is shown in the dot product datapath. \par The broadcasted write turns off the compute array earlier, thereby improving the energy of the system. No weight replication for a bit-serial PIM approach improves the PIM compute density to ((m*n)/n = m) along with reduced area requirements.
\begin{comment}
\begin{figure*}[t!]
\centering
\includegraphics[width=\linewidth]{Aggregation_unweighted_undirected.png}
\vspace{-2em}
\caption{\textbf{a) \underline{UWC engine}: Performs aggregation of unweighted graphs by (i) Reading the adjacency matrix (encoding in Compressed Sparse Row Format) and NodeProc register (indicating the node processed by combination) to fill the update index register to indicate the nodes to be aggregated. (ii) Node 5 (self loop), Node 1 and 3 is aggregated with incoming combination vector for node 5, using a row of adders by "broadcasting" the vector to the aggregation array, enabling "compute-as-soon-as-ready" approach. b) \underline{Aggregation amortization}: Dot product PIM for combination/read of adjacency matrix of n\textsuperscript{th}/(n-1)\textsuperscript{th} node overlap; Filling of combination/aggregation array for n\textsuperscript{th}/(n-1)\textsuperscript{th} node overlap }}

% b/w combination and aggregation at steady state %ggregation arrays are shown on the left for 3*3 weight matrix with 5 nodes in the unweighted, undirected graph and 3 features per node. Update Index and Node Proc are used to track the processed node, indices to be updated in the combination array. Post aggregation using a row of adders b) Overlap b/w combination and aggregation at steady state }}
\label{fig:Aggregation_UWC}
\vspace{-1em}
\end{figure*}
\end{comment}
\begin{figure*}[t!]
\centering
\includegraphics[width=\linewidth]{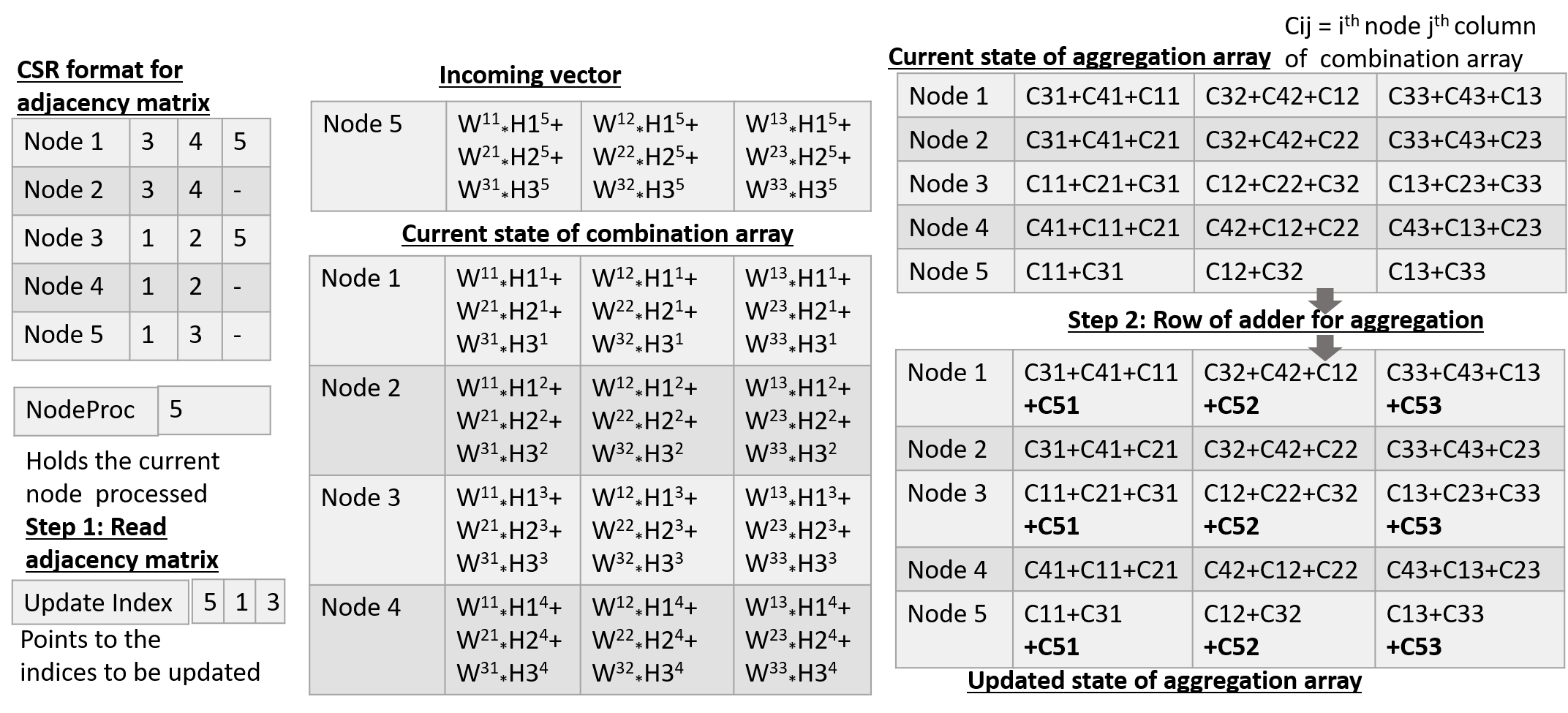}
\vspace{-2em}
\caption{\textbf{Incoming graphs are mapped onto different engines based on graph-connectivity \underline{(graph-aware)} and read-out of adjacency matrix (stored in Compressed Sparse Row Format) to eliminate unnecessary compute \underline{(sparsity-aware)}.  \underline{UWC engine}: Aggregation of unweighted graphs by reading the adjacency matrix  and NodeProc register (indicating the node being processed by combination) to fill the update index register in step 1 and updating the aggregation array using adders in step 2. (ii) Node 5 (self-loop), Node 1 and 3 is aggregated with incoming combination vector for node 5, by \underline{"broadcasting"} the vector (C51) to the aggregation array. The aggregation array is updated immediately with C51 once the combination result for a particular node is available, enabling \underline{"compute-as-soon-as-ready"} approach}}
\label{fig:Aggregation_WC}
\vspace{-1em}
\end{figure*}
\subsection{NEM-C3: Scalable bit-serial PIM with pre-compute}
\par NEM-C2 introduces an ECT data path, which can be mitigated by leveraging the fact that all H bits are available from the storage array read. NEM-C3 tries to eliminate ECT overhead and enhance performance, by virtue of increased pre-computation. 
\par Similar to NEM-C2, multi-bit weights are stored in a single compute array row, with H mapped onto RWL in a bit-serial fashion. Bank/tile-level parallelism for parallelism across j/i-dimension in Hj\textsuperscript{i}\textsubscript{n} is leveraged, without requiring data replication (as shown in Fig.\ref{fig:ETC_CIM}).
\par During the computation of dot products, NEM-C2 awaits the first occurrence of '1' in the H element before terminating the compute. In contrast, in NEM-C3, the computation terminates even earlier by pre-computing dot products corresponding to both '1' and '0' H-bits. This approach obviates the need for the ECT datapath, while still utilizing the existing PIM and dot product datapaths to implement NEM-C3.  The first 2 steps of the PIM data path remain the same as NEM-C2. However, in step 3, both the dot products are broadcasted to fill the partial-products array based on the H bit (shown as blue AND in Fig. \ref{fig:ETC_CIM}), eliminating the overhead of ECT. This helps reduce 3 write ports (3:1 MUX) on the partial products array to a simple logical AND operation between the bank read and the bit for the row in partial products array (H1\textsuperscript{1}\textsubscript{2} for 2\textsuperscript{nd} row).  This AND operation effectively signifies whether the array should be populated with 'zero' or the actual H value, depending on whether the H-bit mapped onto the RWL is '0' or '1'. The dot-product datapath itself remains unchanged from NEM-C3.
\par This results in overall area/power reduction, improved compute density due to ECT elimination, and reduction in write ports. Furthermore, the precompute strategy helps with achieving improved performance and energy, as the compute can be terminated earlier than NEM-C2.

\begin{figure}[t]
\centering
\includegraphics[width=9cm]{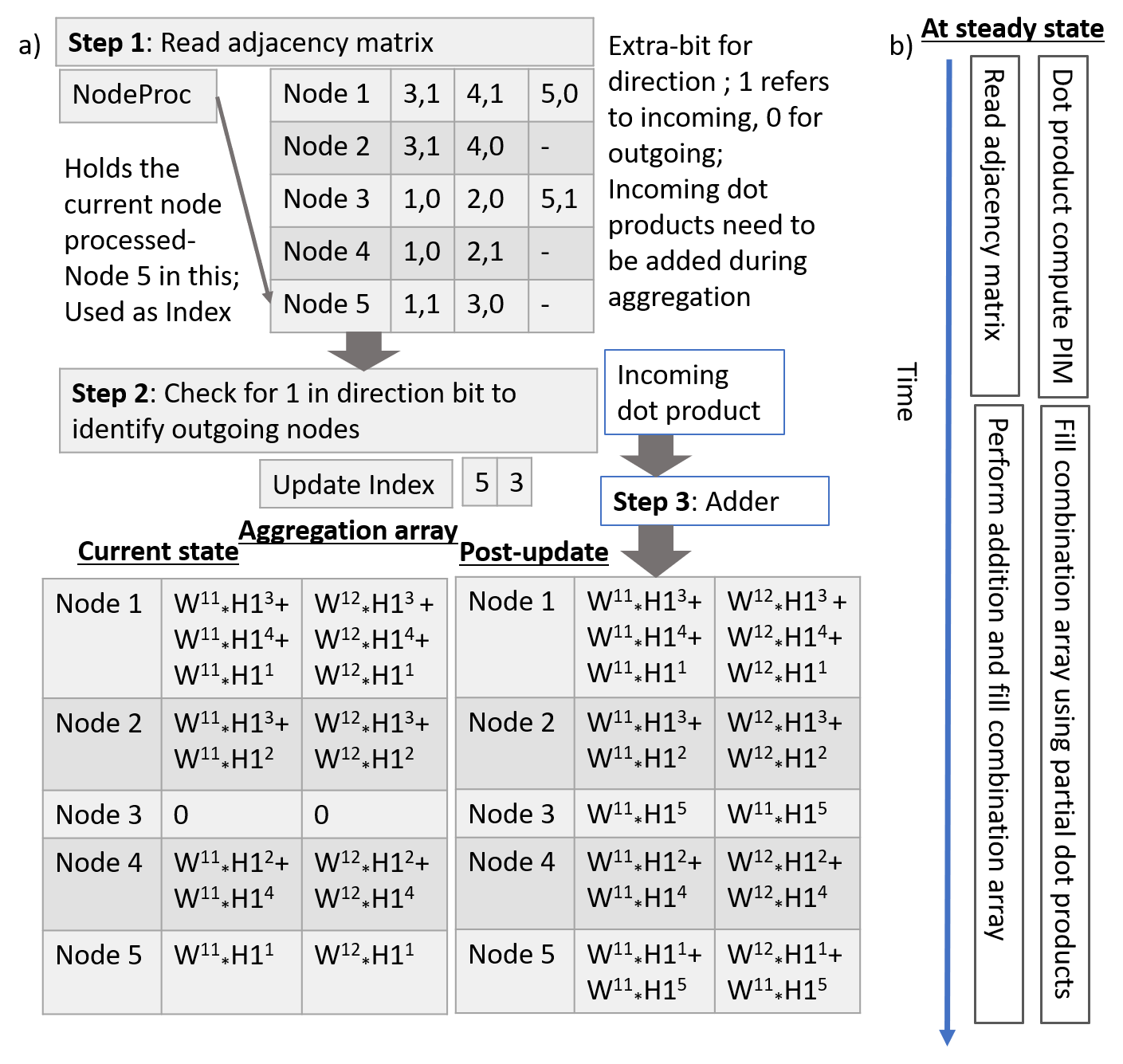}
\vspace{-1em}
\caption{\textbf{a) \underline{UWC engine}: Aggregation for an unweighted, directed graph begins with reading the adjacency vector corresponding to Node Proc in Step 1, identifying outgoing nodes in step 2, and storing in Update Index register, using adders to aggregate the incoming combination vector onto the nodes in Update Index register in step 3. Each adjacency matrix element is of the form (i,j), where i/j represents the neighboring node/direction of interaction with the neighbor b) Timing diagram to show that dot product PIM for combination/read of the adjacency matrix of (n+1)\textsuperscript{th}/(n)\textsuperscript{th} node overlap; Filling of combination/aggregation array for (n+1)\textsuperscript{th}/(n)\textsuperscript{th} node overlap, \underline{hiding aggregation latency}}}
\label{fig:Aggregation_UWC_unweighted_undirected}
\vspace{-1em}
\end{figure}

\begin{figure}[t]
\centering
\includegraphics[width=10cm]{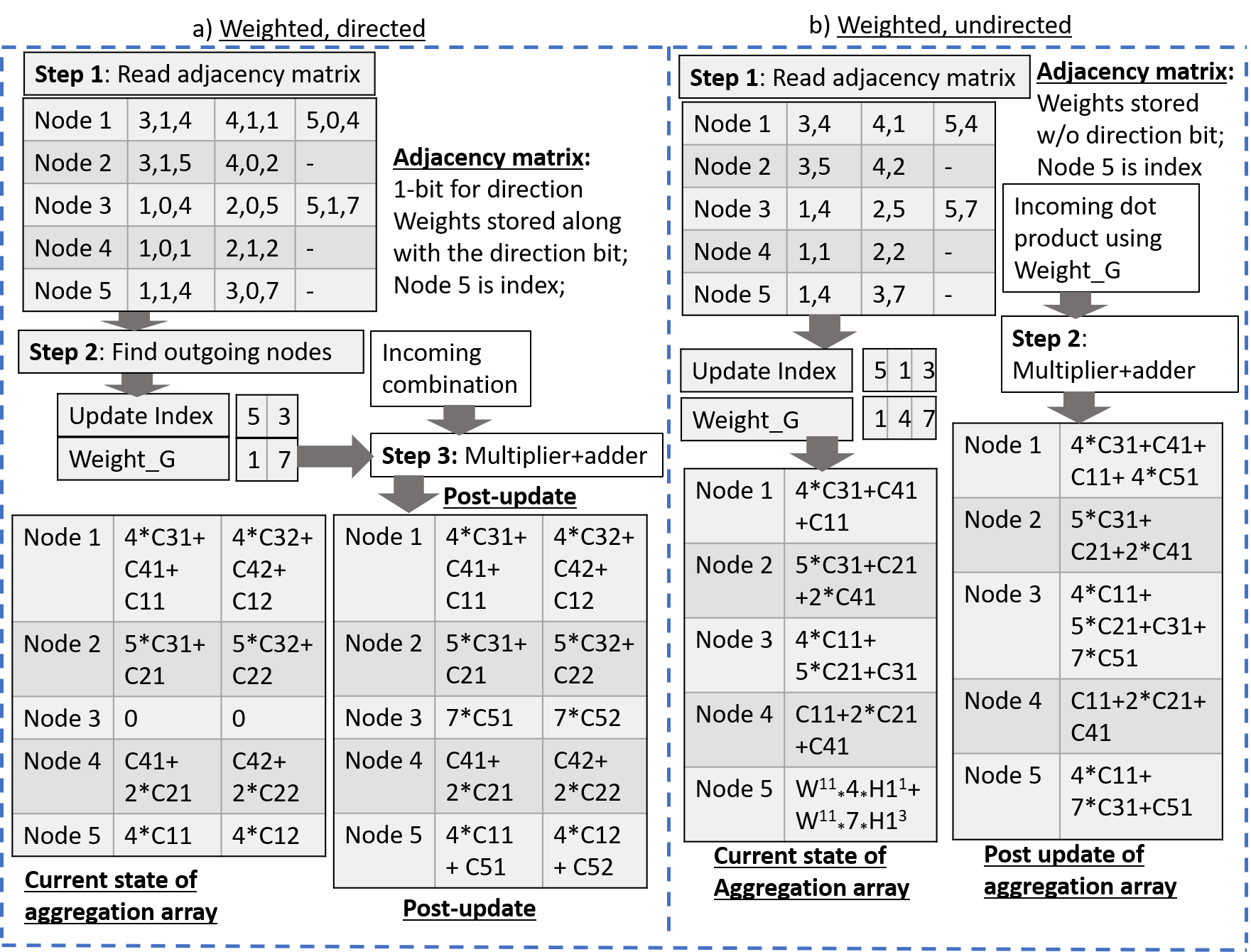}
\vspace{-1em}
\caption{\textbf{a) Weighted, directed aggregation, with adjacency matrix storing the weights of graphs and the direction in the case of directed graphs. The direction is read out in step 1 to check for outgoing nodes in step 2 and aggregation with the incoming combination vector is achieved using near-memory multipliers and adders in step 3 b) Weighted, undirected aggregation follows the same datapath as the directed one, but without the notion of direction, making it a 2-step operation}}
\label{fig:Aggregation_weighted}
\vspace{-1em}
\end{figure}
\begin{comment}
\begin{figure}[t]
\centering
\includegraphics[width=6.5cm\textwidth]{Aggregation_weighted_undirected.png}

\caption{\textbf{Weighted, undirected graph aggregation, with adjacency matrix storing the weights of graphs, using near-memory multipliers and adders to perform aggregation with the incoming combination vector.}}
\label{fig:Aggregation_weighted_undirected}

\end{figure}
\end{comment}

\section{GRAPH AND SPARSITY-AWARE NEAR MEMORY AGGREGATION}
Aggregation involves a two-step procedure, with the first being the multiplication of the resultant combination output with the adjacency (A) matrix, and the second being the multiplication of the outcome from the first step with the D matrix (assuming GCN for simplicity of explanation). In the context of the first step, which is carried out in the WC/UWC engine, the computation is both graph-aware and sparsity-aware. The graph structures are intelligently optimized and distributed across different engines based on connectivity patterns, while the computation is made aware of matrix sparsity to mitigate unnecessary operations involving the A matrix. Additionally, we introduce two approaches: the "compute-as-soon-as-ready" (CAR) strategy, and "broadcast" technique. These techniques help overlap aggregation with combination, enabling resource reuse and concurrent processing. Moving on to the second step, this operation is executed within the D generator, and the process adopts a "sparsity-aware" approach. This technique strategically reduces the number of computations by accounting for the matrix's sparse characteristics. %Overall, the aggregation data path in the proposed accelerator takes into account the nature of the graph that is given as an input and optimizes it accordingly.

\subsection {Graph and sparsity-aware UWC engine for Unweighted graphs}
\par While Graph Convolutional Networks (GCNs) find application across various graph types, their most frequent utilization occurs with unweighted graphs, notably in citation networks like Citeseer and Cora. We enhance the efficiency of these specific graphs by skillfully mapping them onto the Unweighted Control (UWC) engine. This mapping capitalizes on inherent graph patterns and reusability, effectively minimizing the hardware demands.

Several key challenges are evident from prior approaches:
(i) In the context of ReRAM-PIM-based aggregation, the Processing-in-Memory (PIM) array computes partial dot products, necessitating an additional array (referred to as the "aggregation array") to store these partial dot products before they are collectively accumulated. This requires increased area.
(ii) ReFLIP employs the same PIM array for both combination and aggregation, leading to a situation wherein aggregation only commences after the combination process concludes.
(iii) The process of writing the resultant combination array onto PIM array before aggregation introduces an added energy cost. Contrasting this, NEM-GNN adopts a distinct strategy. It does away with the requirement for a PIM array dedicated to aggregation, instead using the "aggregation array" to initiate aggregation as soon as a fresh entry is introduced into the combination array. This facilitates the concurrent execution of aggregation and combination, resulting in improved performance, while eliminating the need for power and area overhead associated with PIM array write and read operations. %(iii) In the case of unweighted graphs mapped onto ReFLIP for aggregation, wherein the resultant combination matrix is mapped onto PIM array, dot product with adjacency vector involves converting the incoming adjacency element ('1' or '0') into analog value using DAC before mapping onto RWL and (a) turning ON multiple WLs at the same time if the DAC output is of finite value (b) Not turn ON any WLs if DAC output is zero, to perform MAC operation. However, in both (a) and (b), the power dissipated in DAC would be in-efficient as it is converting 1-bit digital value and the unnecessarily extra computations when multiplying with '0' leads to higher power. 

\par For aggregation in undirected, unweighted graphs, we begin by reading the adjacency vector (stored in compressed sparse row format (CSR)) for the node undergoing combination (indicated by the "NodeProc" register) in Fig.\ref{fig:Aggregation_WC}. In step 1, reading the adjacency vector highlights nodes that share an adjacency with the specific node in focus.
The identified adjacent nodes, along with the "NodeProc" information, are stored within the Update Index register to serve as aggregation candidates. The aggregation array, with indexing based on the Update Index register content, undergoes immediate updates. This update involves the accumulation of the existing aggregation array values with the incoming combination vector through the utilization of a set of adders. This approach, characterized by simultaneous computation whenever data is ready, is known as the "Compute-as-soon-as-ready" strategy. The incoming combination vector is broadcasted to the aggregation candidates, allowing them to accumulate the combination vector concurrently. This mechanism facilitates "aggregation-vector" level parallelism. Fig. \ref{fig:Aggregation_WC} shows that node 5 is the NodeProc by the combination data path. We identify that node 5 is connected to nodes 1 and 3 from the read-out of the adjacency matrix. Therefore, the "Update Index" register shows that nodes 5 (for self-loop), 1, and 3 are the aggregation candidates. Hence, the aggregation array entries of nodes 5, 1, and 3 are added with the incoming combination vector "C51, C52, C53" in step 2. 
 \par For the aggregation of directed, unweighted graphs, the adjacency matrix has an extra bit that shows the direction of interaction with the neighbors.  '0'/'1' indicates an outgoing/incoming edge from/to a node.  This is done so that only nodes having an outgoing edge from "NodeProc" are aggregated. Post-read-out of the adjacency matrix in step 1, CAM with the direction bit for the NodeProc is done to identify nodes that are to be aggregated in step 2 in Fig.\ref{fig:Aggregation_UWC_unweighted_undirected}a). We utilize "CAR" and "broadcast" to improve performance in NEM-C3 as well. When NodeProc = 5, since node 5 has an outgoing edge to node 3 alone, the combination vector for node 5 is added with entries corresponding to nodes 5 and 3 of the aggregation array, using adders in step 3.
 \begin{comment}
 \begin{figure}[t]
\centering
\includegraphics[width=8.9cm\textwidth]{D_gen.png}
\vspace{-2em}
\caption{\textbf{ Degree matrix generator for generating D\textsuperscript{-1} using a sparsity-aware approach of storing non-zero elements and performing element-by-vector multiplication for every row, reducing the number of computations by n and auxiliary control for ReLU and softmax functions }}
\label{fig:D_generator}
\vspace{-1em}
\end{figure}
\end{comment}
\par This approach eliminates redundant multiplications between the incoming combination vector and the adjacency vector when nodes are not adjacent. This refinement makes the design sensitive to sparsity, meaning the aggregation process no longer includes aggregation with a "0" weight for non-neighboring nodes. In Fig.\ref{fig:Aggregation_UWC_unweighted_undirected}b), there's an evident overlap between aggregation and combination. This is achieved by concurrently reading the adjacency vector of the n\textsuperscript{th} node for aggregation while simultaneously calculating the dot product for the (n+1)\textsuperscript{th} node's combination. This overlap continues with the generation of partial products and the filling of the combination array for the (n+1)\textsuperscript{th} node, all while the aggregation process for the n\textsuperscript{th} node is underway. Due to the complete pipelining of both aggregation and combination, the latency introduced by aggregation is effectively hidden, resulting in improved performance. %In terms of performance, because the combination and aggregation overlap almost completely, the cost of performing a separate aggregation (like in ReFLIP) is completely amortized. 

\begin{figure}[t]
\centering
\includegraphics[width=11cm]{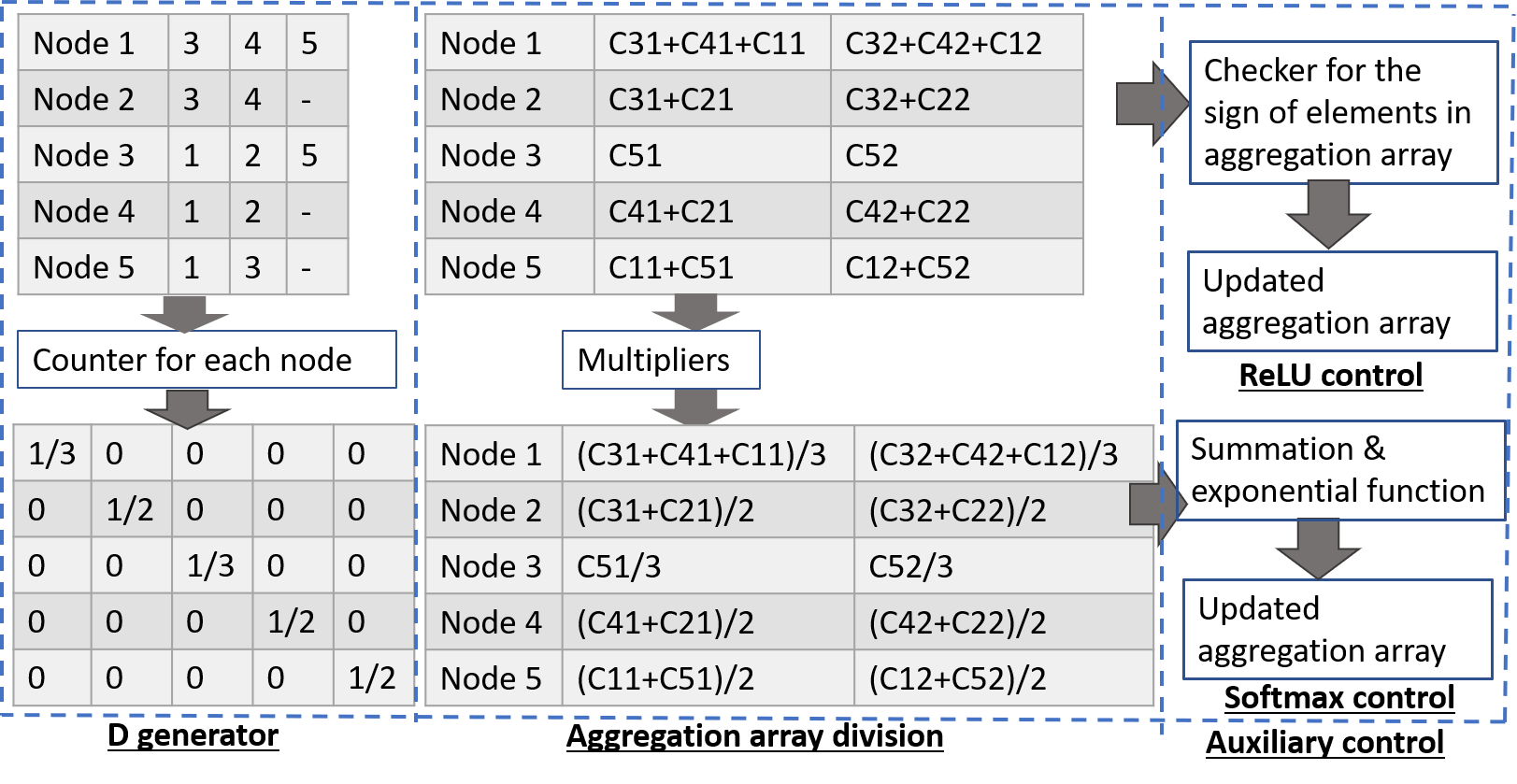}
\vspace{-1em}
\caption{\textbf{ \underline{D-generator and control logic:} Degree matrix generator for generating D\textsuperscript{-1} using a \underline{sparsity-aware} approach that (i) performs element-by-vector (instead of vector-by-vector) multiplication for every row, and (ii) reduces the number of computations/area by a factor of 2n/n. Auxiliary control for ReLU and softmax is shown in the right-most figure.}}% Aggregation for a) unweighted, directed graph using a row of adders with adjacency matrix having a direction bit and a row of adders to aggregate the incoming combination vector \underline{WC engine}: b) Weighted, directed c) Weighted, undirected graphs, with adjacency matrix re-purposed to store the weights of graphs, using a row of multipliers and adders to perform aggregation with the incoming combination vector }}
\label{fig:Sparsity_aware}
\vspace{-1em}
\end{figure}
\subsection{Graph and sparsity-aware WC engine for Weighted graphs}
\par For weighted graphs, the adjacency matrix (A) is re-purposed to store the weight of interaction between 2 adjacent nodes. For compute of directed graphs, post read-out of A in step 1, we find the outgoing nodes of the NodeProc in step 2 to find the update index and weights (indicated as Weight\textunderscore G). Multipliers are used to perform the dot products between the incoming combination vector and the weight corresponding to the edges. Similar to the unweighted graphs, the incoming combination vector is "broadcasted" to accumulate onto the aggregation array corresponding to the update indices using adders in step 3. Fig. \ref{fig:Aggregation_weighted}a) shows that there is an outgoing edge of weight=7 from node 5 to 3, which is multiplied with the incoming combination vector and aggregated onto node 3. For compute of undirected graphs, the adjacency matrix has additional bits (Fig. \ref{fig:Aggregation_weighted}b) for storing the weight of the edge with no peripheral direction detection logic. The adjacency matrix is read in Step 1, to identify the adjacent nodes, which is followed by the "broadcasted" incoming combination vector using "CAR" approach in step 2.  Apart from UWC engine's advantages, WC engine consumes less power than ReFLIP, due to absence of power-hungry DACs/ADCs.

\subsection{Sparsity-aware D generator and control logic}
\par The key observation is that D\textsuperscript{-1} is sparse, and that multiplication of D\textsuperscript{-1} with aggregation array results in unnecessary computations. Therefore, we propose a \textbf{sparsity-aware approach} that consists of two main facets: (a) The approach involves exclusively storing the non-zero elements, specifically the diagonal elements, of the degree matrix (D\textsuperscript{-1}). This curtails the required area by a factor of n. (b) Multiplying the aggregation array with D\textsuperscript{-1} is realized as element-by-vector multiplication for every row, reducing the number of computations by a factor of 2n, for D\textsuperscript{-1} of n*n and the aggregation arrays of n*n. This is because, when 2 arrays each of size n*n array are multiplied with each other, the total number of MAC operations would be $\sim$ 2*n*n*n, assuming 1 operation each for multiplication and addition. However, the proposed approach necessitates only n*n multiplication operations without any accumulation operations. This is achieved by capitalizing on the characteristic of multiplication of a diagonal matrix (D\textsuperscript{-1}) with an adjacency matrix.  This involves multiplying each row in the adjacency matrix with the corresponding non-zero element in the same row of D\textsuperscript{-1}, thereby converting a vector-by-vector product to element-by-vector multiplication. Fig.\ref{fig:Sparsity_aware} shows that D\textsuperscript{-1}\textsubscript{11} (1/3) (element) is multiplied with the first row of the aggregation array, instead of the first row of D\textsuperscript{-1} (like in ReFLIP). Furthermore, during compute, the degree matrix is generated in parallel with the first step of aggregation, followed by multipliers that perform multiplication between non-zero D\textsuperscript{-1} element and result of aggregation.  %Fig.\ref{fig:Sparsity_aware} shows that for node 1, there are 3 neighbors thereby, making D\textsubscript{11}\textsuperscript{-1} = 1/3.

\begin{figure}[t]
\centering
\includegraphics[width=\linewidth]{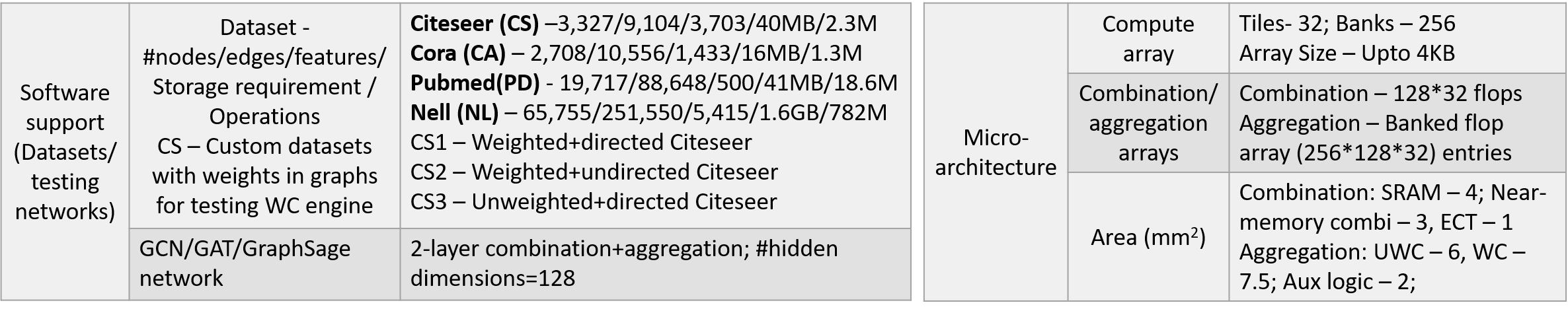}
\vspace{-2em}
\caption{\textbf{\underline{Benchmarks}: Datasets for GNNs, the number of nodes/edges/features in each of them, and the network used for GCN/GAT/GraphSage networks. \underline{Micro-architecture} of NEM-GNN with the additional near-memory logic requiring 2\% of AMD's Zen3 CPU per-core area }}% Aggregation for a) unweighted, directed graph using a row of adders with adjacency matrix having a direction bit and a row of adders to aggregate the incoming combination vector \underline{WC engine}: b) Weighted, directed c) Weighted, undirected graphs, with adjacency matrix re-purposed to store the weights of graphs, using a row of multipliers and adders to perform aggregation with the incoming combination vector }}
\label{fig:Software}
\vspace{-1em}
\end{figure}

\par The advantages of this approach are that a) D matrix is re-used across different layers and b) D-generator is power-gated, once D-matrix is computed the 1\textsuperscript{st} time. This helps achieve improved power/performance while reducing the number of unnecessary sparse computations. The auxiliary \textbf{control logic} consists of (i) ReLU to identify the sign of elements and update the final aggregation array, (ii) softmax control logic  employing exponential/summation to generate classification output. %(i) and (ii) are re-used for generating the attention matrix for GAT. %The final aggregation array is written back to the storage array for processing subsequent layers.

\section{ISA support}
NEM-GNN's reconfigurability (L1/L2 cache used for compute/storage) helps re-purpose CPU instructions like load/store to read/write data onto the L1 cache for compute. Since the L1 cache can operate in 2 modes i.e. normal and compute mode, LCONF instruction configures L1 in compute mode by programming a special purpose register. Additional instructions for performing combination/aggregation are proposed. The instruction MACC Vx, VH, VW carries out an in-memory dot product, followed by near-memory accumulation. This process accomplishes combination between a row of weights (VW) and a feature vector (VH), ultimately storing the outcome in Vx. MACA Vagg, Va, Vx performs near-memory aggregation using the incoming combination vector (Vx) and adjacency vector stored in Va to store the aggregation result in Vagg. 
\section{EXPERIMENTAL METHODOLOGY}
\par \textbf{\underline{Benchmarks}:} A GNN model consisting of 2 graph convolutional layers with 128 hidden dimensions, similar to that of ReFLIP/AWB-GCN \cite{ReFlip}\cite{AWB-GCN} is used as the underlying network to ensure a fair comparison between the proposed and existing designs. The quantization of GNN network weights, feature matrix, and weights of the input graphs (for weighted graphs) is done using PyTorch. The evaluated datasets and their properties are tabulated in Fig.\ref{fig:Software}.

% \usepackage{multirow}
\begin{comment}
   \begin{table}[t]
\caption{Micro-architecture and software support}
\begin{tabular}{|l|l|l|}
\hline
\begin{tabular}[c]{@{}l@{}}Software support\\ (Datasets/ \\ testing network)\end{tabular} & \begin{tabular}[c]{@{}l@{}}GCN dataset \\ (\#nodes/\#edges\\ /feature vector\\ length)\\ CS referring to\\ custom GCNs\end{tabular} & \begin{tabular}[c]{@{}l@{}}Citeseer - 3,327/9,104/3,703\\ Cora - 2,708/10,556/1,433\\ Pubmed - 19,717/88,648/500\\ Nell - 65,755/251,550/5,415\\ CS1 - W+dir Citeseer\\ CS2 - W+undir Citeseer\\ CS3 - UW+dir Citeseer\end{tabular} \\ \cline{2-3}     & GCN network    & \begin{tabular}[c]{@{}l@{}}2-layer GCN + \\ \#Hidden dimensions = 128\end{tabular}   \\ \hline
\begin{tabular}[c]{@{}l@{}}Micro - Architecture \\ (On Chip)\end{tabular}                & SRAM memory                                                                                                                         & \begin{tabular}[c]{@{}l@{}}Tiles - Upto 32\\ Banks - Upto 256\\ Memory Array size - Upto 4KB\\ (32 weights *128 cols *8 bits)\end{tabular}                                                                               \\ \cline{2-3} 
    & Registers    & \begin{tabular}[c]{@{}l@{}}Comb. array - 128*32 flops\\ Aggr. array - Banked flop array\\ (256*128*32) entries/bank\end{tabular}   \\                                               \hline
\end{tabular}

\end{table}
 
\end{comment}

\par \textbf{\underline{Microarchitecture}:}  Compute array, repurposing 8T SRAM L1 cache is organized as 32 tiles/256 banks, matching L1 cache size of Intel Xeon E5-2680 v3. We integrate below mentioned NEM-GNN's near-memory architecture to Intel Xeon E5-2680 v3's architecture (wherever available) to ensure fair comparison, with a banking strategy, as mentioned below.
%Data is mapped onto the array based on the workload, leveraging the parallelism in each of the approaches being evaluated (section IV). 
The combination and aggregation arrays are implemented as registers, with the size of the combination array accommodating 128 W*H dot products each of 32 bits. The aggregation array is organized as banks, with banks differentiated by node number (e.g. Bank0: 0 -255, Bank1: 256-511 nodes). The near-memory logic like shifters/adders are present at 1 per 8 columns, adder reduction/multiplier at 1 per bank.
\par \textbf{\underline{Performance/Power/Area evaluation}:}  Microarchitecture of \textbf{digital components} is specified using System Verilog, and synthesized with FreePDK 45nm \cite{45nmPDK} technology, operating voltage of 1V, clock latency of 2ns to identify the power and area tradeoffs. For the SRAM compute array and its peripheral circuits, power, and area are estimated using Cadence Virtuoso. \textbf{SRAM compute} energy is measured using an SRAM array of size 32*256, with the sense amplifier offset voltage set to 100mV, RBL capacitance of 35fF, as measured from layout extraction with read/write latency of SRAM $\sim$2ns, operating voltage of 1V. We use a custom performance simulator that accounts for the microarchitecture and the workload, to quantify the performance tradeoffs for combination and aggregation. For large-sized graphs not fitting on-chip, the overheads of (i) write latency of compute array is amortized by leveraging separate decoding circuits for write and read logic in 8T SRAM on the periphery, allowing write onto n\textsuperscript{th} row, when m\textsuperscript{th} row is computed, and (ii) data movement from DRAM to storage/compute array is amortized by data-prefetching. The data movement energy is 1pJ/bit ($\sim$ 800x addition energy) \cite{Onur}. 
%and the write/read energy of SRAM is 0.1pJ/bit.  %This simulator is crucial towards estimating the performance of early compute termination (ECT) design, as ECT's performance is very sensitive to the incoming features and weights. 
\begin{comment}
\begin{figure*}[t!]
\centering
\includegraphics[width=\linewidth]{Energy_efficiency.png}
\vspace{-2em}
\caption{\textbf{ \underline{Energy and efficiency comparison}: Energy noramalized to NEM-C3, efficiency measured in terms of throughput/power for GNNs like GCN, GAT and GraphSage. UWC engine is used for aggregation, NEM-C1,NEM-C2 and NEM-C3 used for combination.  }}% Aggregation for a) unweighted, directed graph using a row of adders with adjacency matrix having a direction bit and a row of adders to aggregate the incoming combination vector \underline{WC engine}: b) Weighted, directed c) Weighted, undirected graphs, with adjacency matrix re-purposed to store the weights of graphs, using a row of multipliers and adders to perform aggregation with the incoming combination vector }}
\label{fig:Energy_efficiency}
\vspace{-1em}
\end{figure*}
\end{comment}
\begin{figure*}[t!]
\centering
\includegraphics[width=\linewidth]{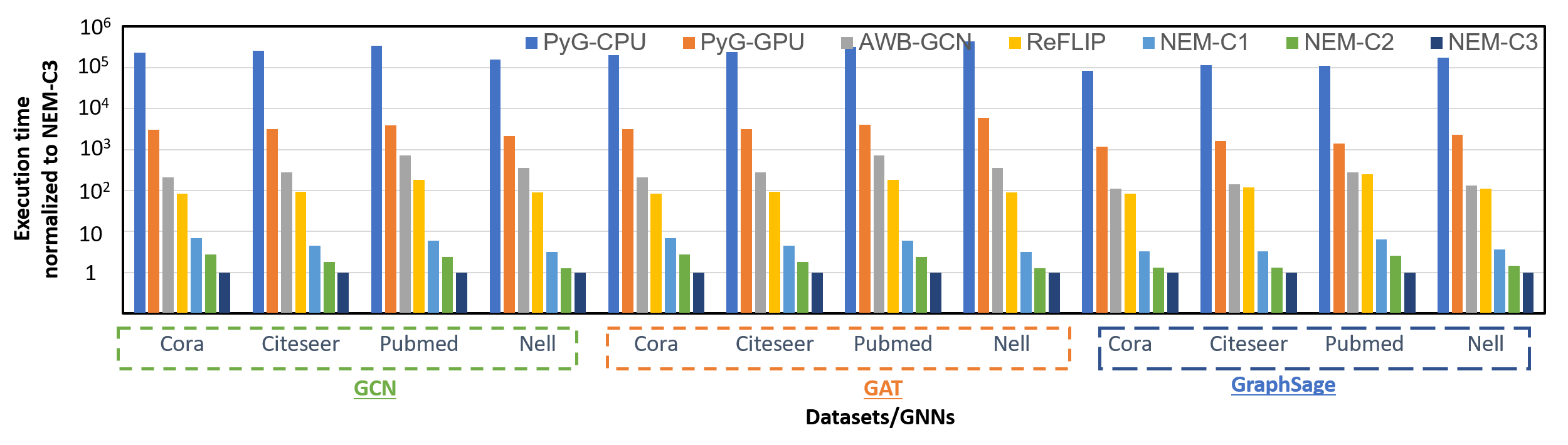}
\vspace{-2em}
\caption{\textbf{ Performance comparison normalized to NEM-C3 for GCN, GAT and GraphSage. UWC engine is used for aggregation, NEM-C1, NEM-C2, and NEM-C3 are used for combination. }}% Aggregation for a) unweighted, directed graph using a row of adders with adjacency matrix having a direction bit and a row of adders to aggregate the incoming combination vector \underline{WC engine}: b) Weighted, directed c) Weighted, undirected graphs, with adjacency matrix re-purposed to store the weights of graphs, using a row of multipliers and adders to perform aggregation with the incoming combination vector }}
\label{fig:Speedup}
\vspace{-1em}
\end{figure*}

\par \textbf{\underline{Comparison Methodology}:} NEM-GNN is compared against PyG-CPU, PyG-GPU (software optimized frameworks of PyG on CPU/GPU), AWB-GCN (non-PIM hardware accelerator) and ReFLIP (PIM hardware accelerator). \textbf{PyG-CPU} is PyG \cite{PyG}, a Python-based GCN-optimized library implementation of the GCN network on Intel Xeon E5-2680 v3, with 12 cores per socket and operating frequency of 2.5GHz, with L1 cache size of 64KB, L2 cache size of 256KB, L3 cache size of 2.5GB, along with DDR4 capacity of 256GB. Similarly, \textbf{PyG-GPU} is implemented on Nvidia Tesla v100, with 64 CUDA cores per streaming multiprocessor (SM) and an operating frequency of 1.5GHz, with 96KB L1 cache per SM, 6MB L2 cache and 16GB HBM2. %The execution times of these designs are recorded for benchmarking the performance.
\textbf{AWB-GCN}'s performance is obtained from its implementation on Intel D5005 FPGA with DRAM capacity of 32GB and 4096 PEs, frequency of 0.3GHz \cite{AWB-GCN}. %taking into account the number of off-chip DRAM accesses and computation latencies. 
The performance of \textbf{ReFLIP} for combination is dependent on (i) the write latency of ReRAM (50.88ns \cite{ReFlip}), (ii) the number of banks and (iii) the number of dot products computed per bank, (16384), limited by the number of DACs/ADCs per bank - 1 per bank in \cite{ReFlip}. %In ReFLIP, the number of dot products computed in a cycle is 16384. %and the total number of dot products for a workload is dependent on the number of hidden dimensions, size of feature element vectors and the number of nodes. 
Aggregation needs to wait until combination completes and is further split into (i) A and (ii) D\textsuperscript{-1} matrix. Multiplication with A and D\textsuperscript{-1} needs to be done serially in a PIM array in GCN/GAT and multiplication with D\textsuperscript{-1} cannot be done in a PIM array for GraphSage because of the exponential function. The overall clock latency is limited by write latency of 50ns, operating voltage of 1V. DRAM prefetching is assumed for large-sized graphs, even in ReFLIP (for fairness). %and the additional write latency of ReRAM is included because of no overlap of combination and aggregation. 
Power is obtained for (i) PyG-CPU using power-stat, (ii) PyG-GPU from Nvidia's system management interface (smi) (iii) Using the same configuration mentioned in \cite{AWB-GCN} for AWB-GCN with rebalancing/distribution smoothing, and (iv) Using the power estimated in \cite{ReFlip} for ReFLIP for DAC/ADC/ReRAM. The additional write costs onto PIM for aggregation, and data movement costs for large-sized graphs are also included. Energy is identified by multiplying the execution time with the power. For NEM-GNN, UWC/WC engine uses unweighted/weighted graphs for aggregation, and NEM-C1,NEM-C2,NEM-C3 are for combination. % power-scaling in \cite{ReFlip} for AWB-GCN, along-with rebalancing/distribution smoothing overheads 

\begin{comment}
\begin{figure}[t]
\centering
\includegraphics[width=8.9cm\textwidth]{Perf_study.png}
\vspace{-2em}
\caption{\textbf{a) Performance comparison normalized to PyG-CPU  b) Average power of the proposed designs c) Energy comparison normalized to GCN\textsuperscript{3} d) Area of the proposed designs e) Maximum throughput on primary axis and energy efficiency normalized to CPU on secondary axis f) Compute density for PIM approaches across different workloads}}
\label{fig:Speedup}
\vspace{-1em}
\end{figure}
\end{comment}

\begin{figure*}[t!]
\centering
\includegraphics[width=\linewidth]{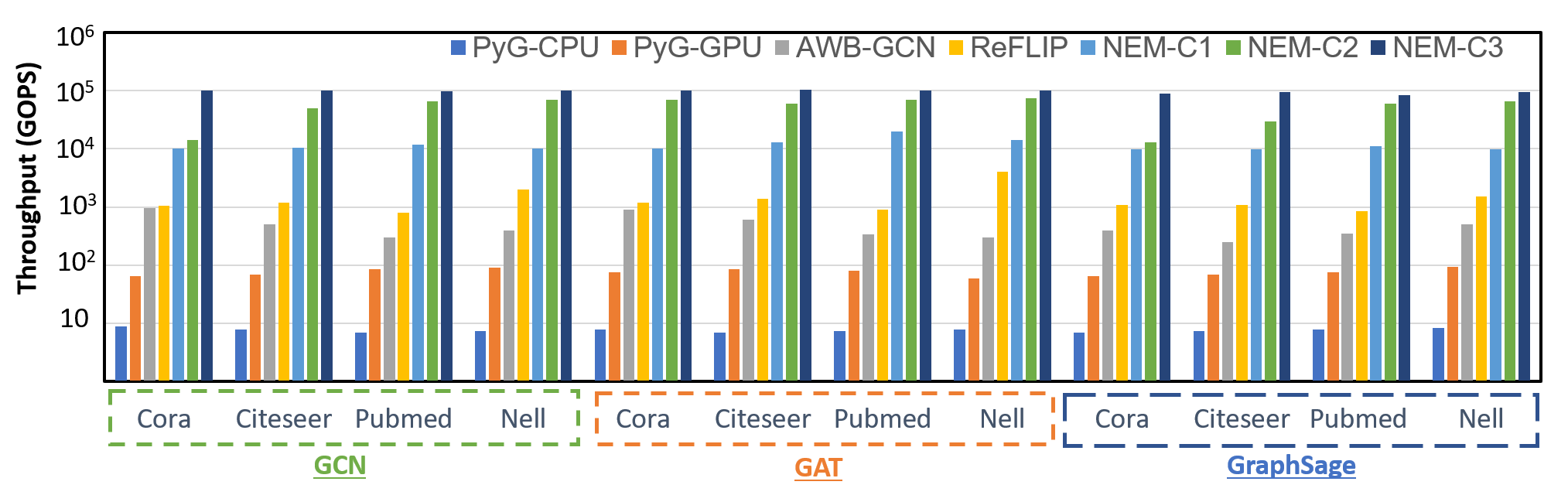}
\vspace{-2em}
\caption{\textbf{ Throughput comparison measured in GOPS for GCN, GAT, and GraphSage. UWC engine is used for aggregation, NEM-C1, NEM-C2, and NEM-C3 are used for combination. }}% Aggregation for a) unweighted, directed graph using a row of adders with adjacency matrix having a direction bit and a row of adders to aggregate the incoming combination vector \underline{WC engine}: b) Weighted, directed c) Weighted, undirected graphs, with adjacency matrix re-purposed to store the weights of graphs, using a row of multipliers and adders to perform aggregation with the incoming combination vector }}
\label{fig:Throughput}
\vspace{-1em}
\end{figure*}
\section{RESULTS}
\textbf{\underline{Performance}} of NEM-GNN (Fig.\ref{fig:Speedup}) is better for \textbf{GCN} than PyG-CPU because the cost of the increased data movement with increased graph size in CPUs is amortized by performing PIM. For PyG-GPU, the irregular memory accesses and limited on-chip memory restrict the performance of the GPUs for larger workloads. For smaller workloads, the ineffective utilization of GPUs due to sparsity limits the performance of GPUs. This is reflected by the increased speedups, as high as $\sim$10\textsuperscript{4} for Pubmed (large graph) and  $\sim$10\textsuperscript{3} for Cora (small graph). To summarize, concerning CPUs/GPUs, apart from in/near-memory compute, early-compute termination/pre-compute strategy, along with hiding latency using CAR and broadcast approaches for aggregation, helps achieve improved performance for NEM-GNN. For AWB-GCN, performance is limited by stalls arising from dynamic adjustment of workloads among 4096 PEs (by performing runtime optimization), and growth in the number of off-chip memory accesses. Furthermore, the multi-stage network (omega) coupled with the control logic necessary for distribution smoothing leads to performance bottlenecks, as it limits the parallelism obtainable across PEs. In NEM-GNN, PIM enables fast memory accesses with high parallelism across all columns/banks, without additional control logic for distribution smoothing, resulting in a speedup of 10\textsuperscript{4}, compared to AWB-GCN in Pubmed. In ReFLIP, the major performance limiters are: (i) 1 DAC/ADC per bank limiting the number of dot products per bank to 1 (ii) serialization of combination and aggregation, %because of the structural hazard of using the same PIM array for both combination and aggregation. 
(iii) high access latencies of ReRAM limiting the cycle time for a single dot product compute. The major advantages of NEM-GNN are (i) for combination, the dot product between a row of weights in a bank and a single H element mapped onto the bank, is computed in parallel (ii) the CAR scheme hides the latency of aggregation completely making the overall latency effectively the time taken to perform combination alone (iii) Lower SRAM access latencies, resulting in speedups of $\sim$80x-200x in almost all workloads. Particularly, NEM-GNN performs well when the ratio of the number of H to the number of nodes is low (observed in Pubmed), as the parallelism across the number of Hs/nodes is dependent on the number of banks/tiles. Among NEM-C* designs, NEM-C1's performance is lower due to the limited parallelism across nodes, NEM-C2's performance is limited by the time to compute 1 node's combination array (as this is data dependent) and NEM-C3 combines the advantages of both NEM-C1 and NEM-C2 for better performance. For \textbf{GAT}, the number of computations is lower (assuming a fixed number of neighbors are predetermined), and NEM-GNN offers better performance($\sim$75-140x) than ReFLIP. For \textbf{GraphSage}, NEM-GNN achieves speedups of $\sim$230x.

\par \textbf{\underline{Throughput}} (Fig.\ref{fig:Speedup}) captures the scalability of different designs. 
PyG-CPU is limited by the number of execution units capable of performing dot products. PyG-GPU is limited by the ability to handle sparse data. AWB-GCN is limited by the number of processing elements and workload rebalancing. ReFLIP's scalability is limited by the number of DAC/ADCs per bank, serial nature of combination and aggregation, and data movement onto the PIM array for initiating aggregation. NEM-GNN overcomes these by performing bit-serial computation, complete use of available throughput from the different columns in a bank (parallel dot product between a row of weights and incoming H element), pre-compute/ECT, parallel combination and aggregation, sparsity-aware compute leading to $\sim$80-300x times higher throughput than ReFLIP.

\begin{figure*}[t!]
\centering
\includegraphics[width=\linewidth]{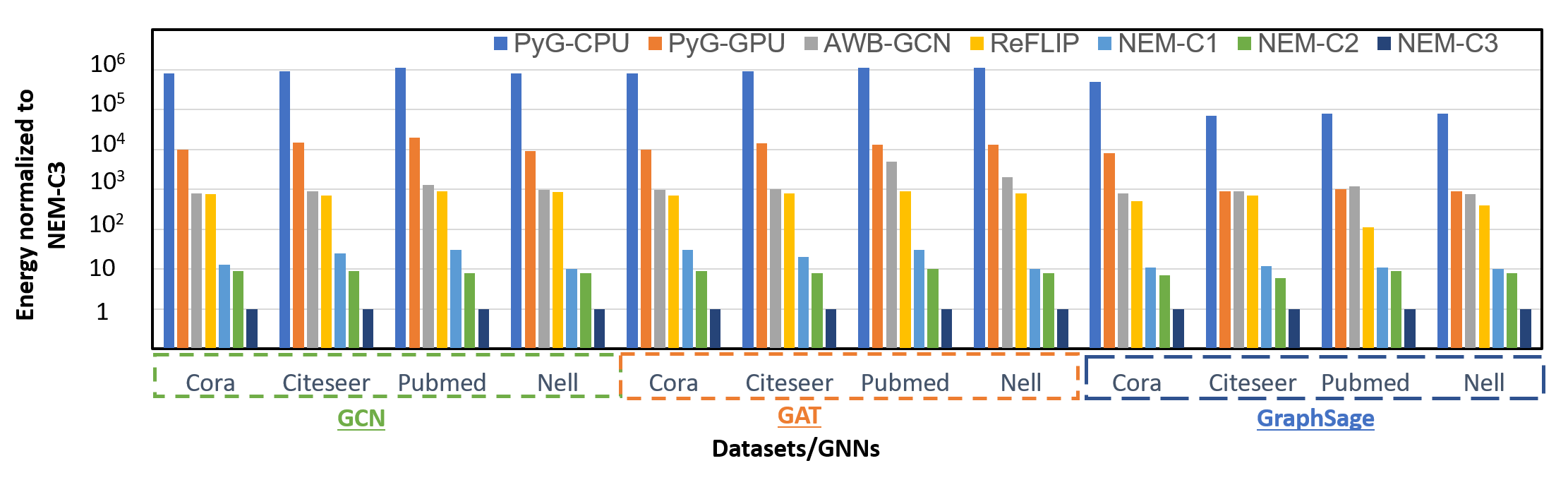}
\vspace{-2em}
\caption{\textbf{ Energy comparison for GCN, GAT and GraphSage. UWC engine is used for aggregation, NEM-C1, NEM-C2, and NEM-C3 are used for combination. }}% Aggregation for a) unweighted, directed graph using a row of adders with adjacency matrix having a direction bit and a row of adders to aggregate the incoming combination vector \underline{WC engine}: b) Weighted, directed c) Weighted, undirected graphs, with adjacency matrix re-purposed to store the weights of graphs, using a row of multipliers and adders to perform aggregation with the incoming combination vector }}
\label{fig:Energy}
\vspace{-1em}
\end{figure*}
\begin{figure*}[t!]
\centering
\includegraphics[width=\linewidth]{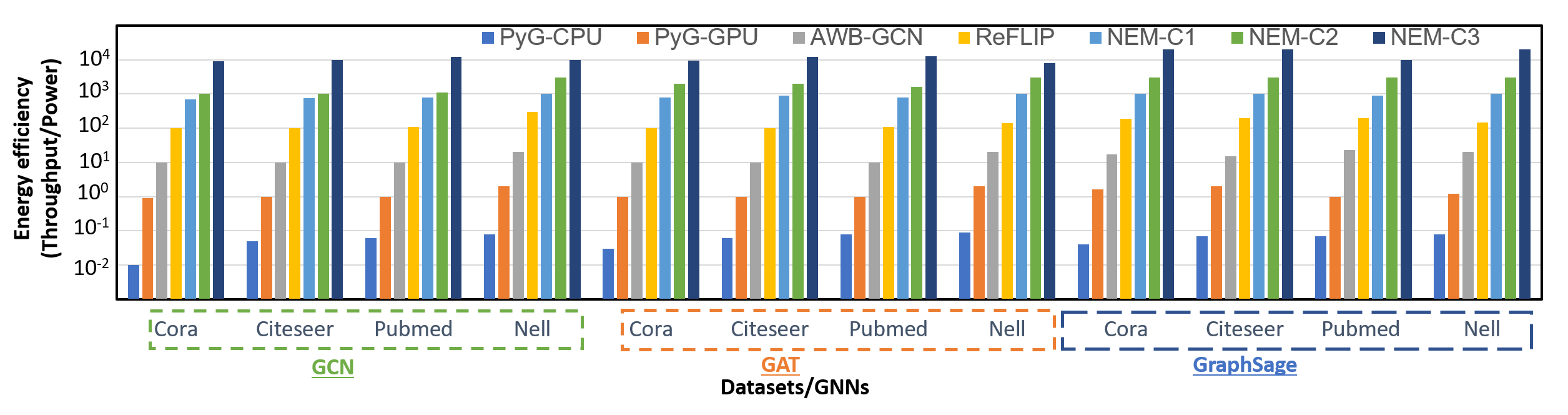}
\vspace{-2em}
\caption{\textbf{ Energy efficiency comparison for GCN, GAT and GraphSage. UWC engine is used for aggregation, NEM-C1, NEM-C2, and NEM-C3 are used for combination. }}% Aggregation for a) unweighted, directed graph using a row of adders with adjacency matrix having a direction bit and a row of adders to aggregate the incoming combination vector \underline{WC engine}: b) Weighted, directed c) Weighted, undirected graphs, with adjacency matrix re-purposed to store the weights of graphs, using a row of multipliers and adders to perform aggregation with the incoming combination vector }}
\label{fig:Energy_efficiency}
\vspace{-1em}
\end{figure*}
\par \textbf{\underline{Energy:}} The overall power of NEM-GNN is divided into the power for (i) Combination control logic (ECT data path, dot product data path)/arrays (combination/dot-product array), (ii) UWC engine (adder)/arrays (adjacency matrix, aggregation array), (iii) auxiliary control logic, (iv) SRAM power across banks/tiles (v) Dot product array and multipliers in WC engine and (vi) ECT data path. The overall estimated power from (i) is $\sim$3W, (ii) is $\sim$6W and (iii) is $\sim$1W and (iv) is $\sim$2W. (v) and (vi) in NEM-C2 costs additional 1.5W power (Fig.\ref{fig:Speedup})). Among NEM-GNN designs, NEM-C3 shows the least amount of energy (Fig.\ref{fig:Speedup}) because the improved performance compensates for the power overhead. NEM-C2 has a slightly higher energy because of the lower performance and increased power from the additional ECT data path, as opposed to NEM-C3. Although NEM-C1 has the least power, the overall energy is higher because the decreased performance overcompensates the lower power. In comparison to ReFLIP, NEM-GNN has the following advantages: (i) No power-hungry DAC/ADC requirements (ii) Lower write/read voltages for SRAM than ReRAM (iii) No additional write required to store back into the compute array post combination resulting in energy improvements of 10\textsuperscript{2-3} in NEM-GNN (\textbf{GCN}). %In comparison to AWB-GCN, there is (i) No movement of data from off-chip memory to compute units, reducing power overhead (ii) No requirement of re-balancing of workloads (causing $\sim$ 40-50\% of power overhead in AWB-GCN), resulting in energy improvements. 
PyG-CPU, PyG-GPU, and AWB-GCN suffer from irregular memory accesses and frequent data movement, costing energy. Furthermore, there is no re-balancing of workloads like in AWB-GCN. In \textbf{GAT}, the energy improvement is higher because of the low power requirement in NEM-GNN, from lesser computations, along with the advantage from improved performance. In \textbf{GraphSage}, the power dissipated is higher (increased number of computations) with similar latency (latency of M matrix generation is completely hidden).

\par \textbf{\underline{Efficiency}} is calculated by dividing the throughput by power (Fig.\ref{fig:Speedup}). CPUs have limited on-chip memory and throughput, with the least efficiency. GPUs have higher throughput with many on-chip processing cores and improved on-chip memory capacity, indicating better energy efficiency as opposed to CPUs. AWB-GCN's decreased throughput is compensated by better energy, improving the efficiency over CPUs. ReFLIP performs PIM based compute with a higher degree of parallelism, improving throughput and energy. NEM-GNN has the highest amount of parallelism leading to increased throughput and decreased energy. Bit-serial PIM combination with pre-compute, and near-memory aggregation results in $\sim$850-1134x improvement over ReFLIP.

\begin{figure*}[t!]
\centering
\includegraphics[width=\linewidth]{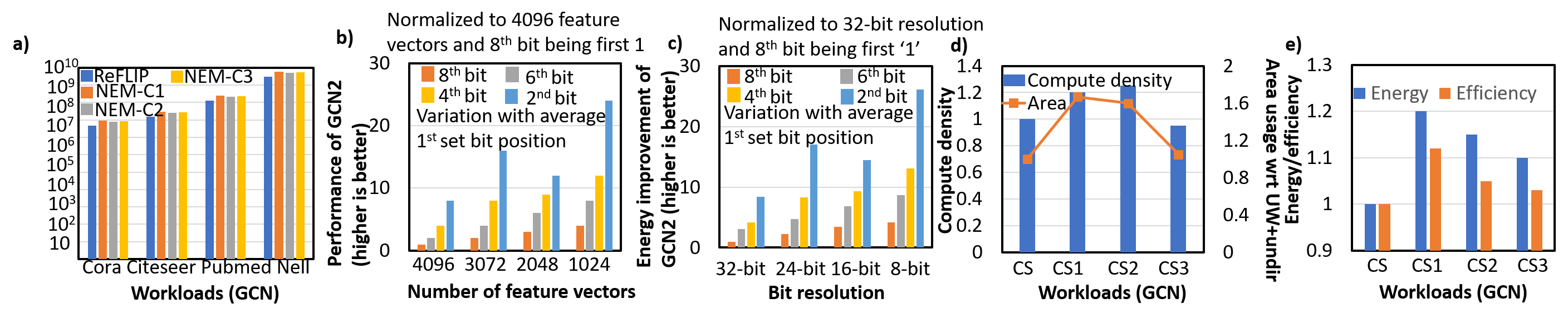}
\vspace{-2em}
\caption{\textbf{a)  \underline{Compute density} comparison across PIM designs b) \underline{NEM-C2 performance} variation with number of Hs c) \underline{NEM-C2 energy} variation with bit resolution, average bit-position for first '1' d) \underline{Compute density, area} for CS1, CS2 and CS3 e) \underline{Energy, efficiency} for CS1, CS2, and CS3}}
\label{fig:DSE}
\vspace{-1em}
\end{figure*}

\par \textbf{\underline{Utilization:}} Fig.\ref{fig:DSE}a) tracks the resource utilization efficiency in terms of the compute density, which is equal to the number of computations (dot products) per unit area.  The overall additional near-memory logic area required per CPU core is tabulated in Fig.\ref{fig:Software} and sums to 0.47mm\textsuperscript{2}, which is 2\% of AMDs Zen3-CPU area. UWC and WC engines share most of their data paths, thereby saving area. NEM-C1+UWC/WC engine has an area requirement of 0.27mm\textsuperscript{2}. NEM-C2+UWC/WC engine has an additional 0.2mm\textsuperscript{2} due to the presence of ECT data path, compared to NEM-C1 based design. NEM-C3 has negligible area increase due to the presence of extra AND gate, compared to NEM-C1 based design. The compute density is $\sim$7-8x that of ReFLIP, due to the elimination of bulky DACs/ADCs, no data replication, and sparsity-aware compute. %and compact compute array size. 
\par{\textbf{\underline{Design space exploration}:}} %It is important to understand that NEM-C2 is dependent on the incoming data, (the position at which first '1' is observed on an average). 
The performance of NEM-C2 varies roughly linearly with (i) the average position of the first '1' of the incoming H vector, (ii) the number of Hs, as they determine the average time for combination per node and the required number of dot products (Fig.\ref{fig:DSE}b). For energy (Fig.\ref{fig:DSE}c), (i) lower bit-resolution causes lesser RBLs to discharge implying less power (for a fixed number of nodes/Hs) (ii) a decrease in the number of bits to obtain first '1' implying better performance. Both these factors combined show less energy for low resolution (e.g.8-bit over 16-bit) and decreased number of bits to obtain the first '1' (e.g.2-bit over 4-bit). The area (Fig.\ref{fig:DSE}d) required for evaluating CS1, CS2, and CS3 is more than CS, because of the additional multipliers for processing weighted edges, and the associated control logic for restricting the computation to outgoing edges. Compute density is the highest for evaluating CS2, because increase in area is compensated by increased number of computations. Fig.\ref{fig:DSE}e) shows that the energy (Power*Performance) for CS1 is the highest, as there is a power increase from the multiplier and control logic (even when MAC for weighted edges is performed in a single clock cycle). The energy efficiency is the highest for CS1, as the energy increase is compensated by the increased number of computations.
 \section{Additional comparison results}
 In this section, we compare NEM-GNN's design, utilizing NEM-C3 for combination and UWC engine for aggregation, against other ReRAM-based PIM designs like PIM-GCN \cite{PIM_GCN}, Challapalle et al. \cite{PIM_GCN_2}, FlowGNN (state-of-the-art Von-Neumann accelerator), and PEDAL, focusing on GCN and GraphSage performance/energy metrics. As absolute throughput values aren't provided, a direct energy efficiency/throughput comparison isn't feasible, and GAT results aren't included. 
 \begin{figure*}[t!]
\centering
\includegraphics[width=\linewidth]{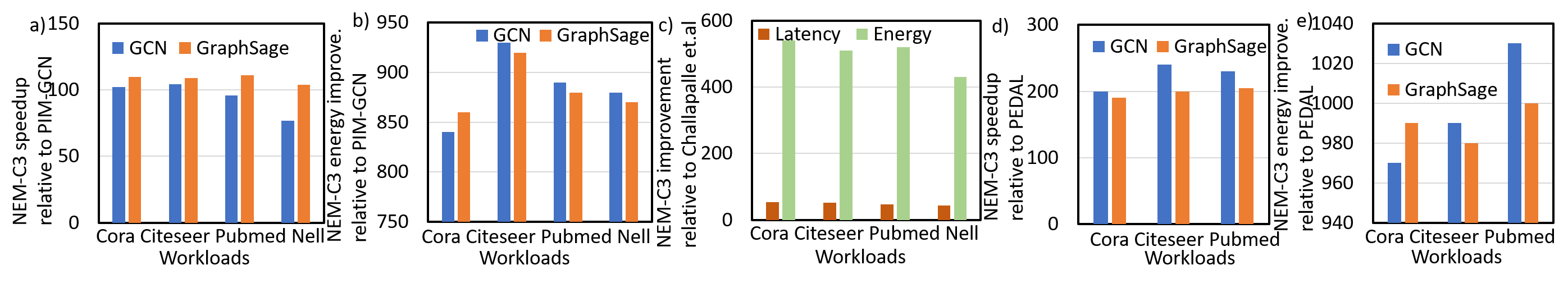}
\vspace{-2em}
\caption{\textbf{a) Performance/b) energy improvement of NEM-C3 relative to PIM-GCN, c) Speedup/energy improvement relative to Challapalle et.al, d) Speedup, e) Energy of NEM-C3 relative to PEDAL}}
\label{fig:ReRAM_PIM}
\vspace{-1em}
\end{figure*}
 \par PIM-GCN demonstrates speedups compared to PyG-CPU running on Intel Xeon E5-2680 v3 (the same hardware used in our simulations). Therefore, we employ the speedup/energy efficiency metrics reported in \cite{PIM_GCN} for comparison. Our focus is limited to GCN and GraphSage for PIM-GCN, as the execution methodology for GAT isn't detailed in the article. While the disadvantages of ReFLIP are applicable to PIM-GCN, NEM-GNN designs achieve higher speedups. However, the achieved speedup is slightly greater compared to ReFLIP in smaller datasets like Cora/Citeseer, mainly because PIM-GCN faces challenges in hiding additional latency for performing CAM to identify neighbors in the scheduling policy, whereas it performs better for larger datasets. This results in speedups of $\sim$ 76x-105x, as depicted in Fig. \ref{fig:ReRAM_PIM}a). Similarly, in terms of energy efficiency, enhancements of $\sim$ 840x-940x are observed for GCN/GraphSage, as shown in Fig. \ref{fig:ReRAM_PIM}b).

 \par Challapalle et.al's provided absolute performance/energy figures guide the comparison. Unlike ReFLIP, this architecture has distinct engines for traversal, combination, and aggregation, potentially enabling faster aggregation. However, NEM-GNN demonstrates speedups of $\sim$ 45x-53x and energy enhancements of $\sim$ 430x-570x, detailed in Fig. \ref{fig:ReRAM_PIM}c). These gains stem mainly from high ReRAM write latency/power and challenges in completely concealing latency due to PIM compute. The combination engine must write results into the aggregation engine before in-memory computation, and all neighboring nodes' combination vectors must be available before initiating aggregation for a node. Otherwise, frequent data transfers between main memory and ReRAM array incur power/latency costs. Moreover, lacking sparsity-aware compute, aside from CSR/CSC data representation, limits potential computational reductions, unlike NEM-GNN's approach, leading to increased energy consumption in the UWC engine. It is important to note that PIM-GCN and Challapalle et.al's modelling does not take into account the additional data movement cost associated with movement from host CPU to ReRAM-based memory array, and that would further improve speedup/energy associated with NEM-GNN.
 
\begin{figure*}[t!]
\centering
\includegraphics[width=\linewidth]{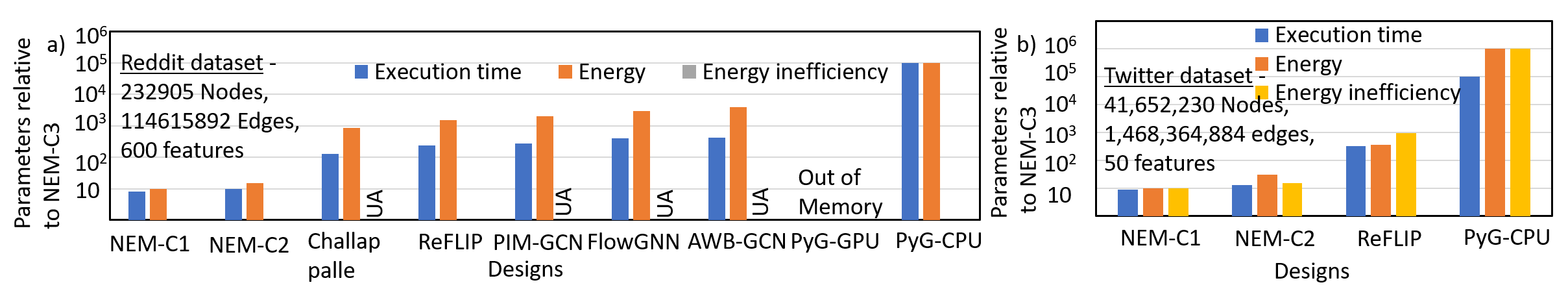}
\vspace{-2em}
\caption{\textbf{a) Execution time/energy requirement/energy inefficiency  of designs relative to NEM-C3 for a) Reddit dataset, b) Twitter dataset. UA means unavailable}}
\label{fig:Reddit_Twitter}
\vspace{-1em}
\end{figure*}

 \par We compare using latency/energy data from FlowGNN. While FlowGNN introduces the capability to compute graphs with edge embeddings, its performance remains nearly on par with previous accelerators for unweighted/undirected graphs when normalized to DSP usage. Hence, the energy efficiency improvements/speedups mirrored in NEM-GNN designs would resemble those seen in AWB-GCN, as depicted in Fig.\ref{fig:Speedup} and Fig.\ref{fig:Energy} for GCN.
 \par For PEDAL, we calculate latencies by multiplying workload cycles by clock frequency and determine energy by multiplying average power with latency. The highest performance across IP-AC, RW-AC, and RW-CA dataflows is reported. We observe $\sim$200-230x/$\sim$190-205x performance improvement in GCN/GraphSage and $\sim$960-1030x/$\sim$980-1000x energy improvement in GCN/GraphSage. These enhancements stem from factors like periodic host-accelerator interaction, constrained throughput from processing engines, and limited opportunities to conceal aggregation latency, owing to inherent Von-Neumann architecture constraints compared to PIM.
\par We compare prior works for Reddit dataset in terms of execution time, energy, energy inefficiency (wherever applicable). Similarly, comparison against Twitter dataset is performed against ReFLIP, PyG-CPU, as other designs have not reported their values. PyG-GPU has out of memory errors for both these datasets. Such large datasets show the scalability potential of NEM-GNN designs. The speedup reason for Von-Neumann architecture designs is the same as mentioned in Sec.9. Comparing NEM-GNN designs with other ReRAM based PIM designs, these are the following advantage:  In large datasets like Reddit or Twitter, because we rely on the combination result to broadcast the data to the aggregation array (instead of aggregation process involving search for combination vectors corresponding to neighboring nodes to be aggregated for every node, like in ReFLIP), and the possibility of combination vector getting broadcasted to more candidates increases. Since, the adjacency matrix is read in parallel to dot-product compute for combination, the broadcast approach can broadcast the combination vector to potentially larger number of aggregation candidates/nodes. Therefore, NEM-GNN is efficient for larger workloads, as well. Furthermore, this is not possible in other PIM designs, as combination vector results need to be written before initiating aggregation. These features enable improved performance/energy/efficiency of NEM-GNN, as shown in Fig.\ref{fig:Reddit_Twitter}. 
\par NEM-GNN can integrate memory-compiled 8T SRAM arrays, improving compute performance and reducing production cycle time. %Memory compiler simulations in TSMC 65nm technology yield: a) libs files for latency/power measurements and b) a lef file for technology/area information comparable to our custom SRAM array. 
Scaled to 65nm \cite{Shanshan_Ising}\cite{SACHI_arxiv}\cite{SACHI}\cite{Ising_review}, our custom array is 1.5 times larger with 1.4x slower read/write latency and 1.6x higher power consumption than compiled memory, attributed to optimized layout reducing BL/WL capacitance. This boosts performance, power efficiency, and reduces footprint, resulting in 1.3x, 1.33x, and 1.35x performance benefits in GCNs, GATs, and GraphSage for NEM-GNN compared to the custom array. Energy efficiency improves up to 1.4x, 1.45x, and 1.47x for GCNs, GATs, and GraphSage, compared to the custom array NEM-GNN.

\section{DISCUSSION}
\subsection{Adaptive Precision and Quantization Support}

Another promising extension for NEM-GNN is the incorporation of adaptive precision execution and hierarchical quantization support. Modern GNN workloads often exhibit varying numerical sensitivity across layers, node types, and aggregation stages. Certain feature transformations require higher precision to preserve convergence accuracy, while aggregation operations can frequently tolerate aggressive quantization. Introducing dynamic precision scaling mechanisms can significantly improve energy efficiency and memory utilization without substantially affecting model quality.

The existing DAC/ADC-less digital execution model of NEM-GNN provides a strong foundation for flexible quantized computation. Since the architecture already avoids expensive analog conversion overheads, extending the datapath to support mixed-precision arithmetic can be achieved with relatively modest modifications. For example, low-precision aggregation units may be used for intermediate feature accumulation, while selective high-precision execution can be reserved for attention computation or normalization layers. Additionally, sparsity-aware quantization schemes may further reduce memory movement by compressing insignificant node features and edge weights.

A runtime-driven precision controller can dynamically adapt numerical formats based on workload behavior, graph sparsity, or layer sensitivity. Such techniques are especially beneficial for large-scale recommendation systems and graph transformers, where memory bandwidth often becomes the dominant bottleneck. Integrating adaptive quantization into NEM-GNN can therefore improve scalability while maintaining the architecture’s core energy-efficient near-memory design philosophy.
\subsection{Runtime Scheduling and Graph Reordering Optimizations}

Efficient runtime scheduling remains a major challenge for irregular graph workloads due to varying node degrees, load imbalance, and unpredictable memory access behavior. An important future extension for NEM-GNN is the development of hardware-software co-designed runtime schedulers that dynamically optimize graph execution order based on workload characteristics. Such runtime systems can improve accelerator utilization, reduce memory contention, and enhance parallelism across near-memory compute units.

The current compute-as-soon-as-ready execution strategy already provides an effective mechanism for latency hiding and asynchronous execution. Building upon this capability, future NEM-GNN designs may integrate graph reordering techniques such as degree-based clustering, locality-aware partitioning, or hot-node prioritization. These methods can improve memory locality and reduce repeated feature fetches during aggregation. Furthermore, dynamic work-stealing schedulers may redistribute computation across idle memory partitions to address load imbalance caused by skewed graph structures.

Another promising direction is integrating predictive scheduling mechanisms that leverage graph topology statistics or machine learning models to anticipate communication bottlenecks. Since graph workloads often exhibit recurring execution patterns, runtime systems can proactively optimize memory placement and aggregation ordering. Combining such runtime intelligence with the existing sparsity-aware NEM-GNN datapath can further improve throughput and scalability for large-scale industrial graph analytics applications.

\subsection{Reliability, Thermal, and Power Delivery Optimizations}

Near-memory accelerators for graph processing introduce highly irregular current demand patterns due to sparse and bursty aggregation behavior. As NEM-GNN scales toward larger graph workloads and distributed memory systems, power delivery and thermal management become increasingly important research challenges. Simultaneous activation of multiple aggregation units can generate localized hotspots and transient voltage droop, potentially impacting timing reliability and long-term device lifetime.

Several architectural characteristics of NEM-GNN make it well suited for adaptive power-aware execution. The sparsity-aware aggregation framework already reduces unnecessary computation activity, indirectly lowering switching power and dynamic current demand. Additionally, the reconfigurable scheduling mechanisms may be extended to throttle computation based on thermal constraints or instantaneous power budgets. For example, aggregation tasks can be spatially distributed across memory partitions to avoid concentrated thermal hotspots while maintaining high throughput.

Future extensions may also incorporate power-aware graph scheduling and thermal balancing algorithms that monitor utilization across near-memory compute regions. Lightweight on-chip telemetry mechanisms can dynamically adjust aggregation intensity, communication frequency, or memory activation patterns to maintain stable operating conditions. Such reliability-aware enhancements are particularly important for large-scale datacenter deployments where sustained graph processing workloads can stress the power delivery network over extended execution periods.
\subsection{Extending NEM-GNN to Broader Sparse Workloads}

Although NEM-GNN is primarily designed for graph neural network acceleration, many of its underlying architectural principles are broadly applicable to a wide range of sparse and irregular workloads. Applications such as sparse matrix--matrix multiplication (SpMM), sampled dense-dense matrix multiplication (SDDMM), PageRank, breadth-first search (BFS), sparse attention mechanisms, and integer linear programming (ILP) all exhibit irregular memory accesses, sparse computation patterns, and significant data movement overheads. Extending NEM-GNN beyond GNN inference therefore represents a promising direction for developing a general-purpose sparse near-memory accelerator.

A major advantage of NEM-GNN lies in its sparsity-aware execution model and compute-as-soon-as-ready scheduling framework. These mechanisms naturally align with sparse linear algebra kernels such as SpMM and SDDMM, where only a subset of matrix entries participate in useful computation. By leveraging the existing sparse aggregation datapaths, NEM-GNN can dynamically skip zero-valued operands and avoid unnecessary memory accesses. Similarly, PageRank and BFS workloads rely heavily on sparse graph traversal and neighborhood propagation, making them suitable candidates for near-memory sparse aggregation and broadcast-based communication.

Several existing architectural components within NEM-GNN can already support these broader workloads with relatively modest modifications. The reconfigurable datapath and broadcast communication infrastructure may be reused for sparse feature dissemination, frontier propagation, and partial result accumulation. Furthermore, the early termination mechanisms originally designed for GNN aggregation can be adapted to sparse iterative algorithms where convergence occurs incrementally. For example, PageRank updates with negligible contribution may be pruned dynamically to reduce computation overhead and improve energy efficiency.

Another promising extension is supporting sparse attention workloads used in transformers and large language models. Sparse attention introduces irregular token dependencies and selective feature interactions that resemble graph aggregation behavior. NEM-GNN’s sparsity-aware execution engine can potentially accelerate top-k attention, block-sparse attention, or locality-aware attention mechanisms by selectively processing only significant token interactions. Such extensions would substantially broaden the applicability of NEM-GNN toward emerging AI workloads beyond graph analytics.

In addition to machine learning applications, the architecture may also accelerate optimization-oriented sparse workloads such as integer linear programming (ILP). ILP solvers often involve sparse constraint matrices, branch-intensive execution, and irregular memory accesses that are poorly suited for conventional GPUs. The near-memory execution model of NEM-GNN can reduce data movement overhead during sparse constraint evaluation and matrix operations. Furthermore, its asynchronous compute \cite{Ising_arxiv} scheduling mechanisms can naturally support irregular branch-and-bound style execution patterns observed in optimization algorithms.

Future work may therefore explore transforming NEM-GNN into a unified sparse near-memory computing framework capable of supporting graph analytics, sparse AI models, optimization kernels, and scientific sparse linear algebra applications \cite{ILP_arxiv}\cite{SPARK}. Such a generalized accelerator would address a growing class of workloads limited not by arithmetic throughput, but by irregular memory movement, sparse data dependencies, and inefficient data orchestration in traditional computing architectures.
\section{CONCLUSION}
We propose a scalable, reconfigurable, DAC, ADC-less, energy-efficient, high-performance GNN accelerator that reconfigures the L1 cache to realize PIM architecture for performing combination and a near memory approach for performing aggregation. For combination, bit serial PIM designs with early compute termination, and pre-compute are proposed to make the design scalable, without requiring data replication and DAC/ADC. For aggregation, graph, and sparsity-aware approaches leveraging the underlying graph connectivity, sparsity combined with "compute-as soon-as-ready" and "broadcast" approaches hide the aggregation latency to improve performance. %Furthermore, we effectively map the GCN algorithm onto hardware, which helps re-use hardware resources providing higher compute density and increased throughput, without the use of DAC/ADC. 
%%%%%%% -- PAPER CONTENT ENDS -- %%%%%%%%

%%%%%%%%% -- BIB STYLE AND FILE -- %%%%%%%%
\bibliographystyle{ACM-Reference-Format}
\bibliography{refs}

\end{document}